\documentclass[12pt]{article}

\usepackage[numbers]{natbib}

\usepackage{varioref}

\IfFileExists{srcltx.sty}{\usepackage[active]{srcltx}}

\usepackage{amssymb,amsmath,bm,array,longtable,colortbl}%
\usepackage{graphicx}%  
\usepackage{slashbox}% Diagonal divider in table cells

         \newcount\hour \newcount\minute \hour=\time \divide \hour by 60
         \minute=\time \count99=\hour \multiply \count99 by -60 \advance
         \minute by \count99  % "\now" prints current time

         \usepackage{xspace,paralist} %
         \usepackage[a4paper,hscale=.72,vscale=.7,nohead]{geometry} %

\usepackage[colorlinks=true,bookmarks=false]{hyperref}

\IfFileExists{hypernat.sty}{\usepackage{hypernat}}

\newcommand{\gev}{\ensuremath{\:\mathrm{GeV}}} %
\newcommand{\kev}{\ensuremath{\:\mathrm{keV}}} %
\newcommand{\ev} {\ensuremath{\:\mathrm{eV}}}   %
\newcommand{\mpc}{\:\mathrm{Mpc}} % Mpc
\newcommand{\dm}{{\textsc{dm}}} % Dark Matter
\newcommand{\numsm}{$\nu$MSM\xspace} % nuMSM

\newcommand{\CL}{\ensuremath{\mathcal{L}}}

\newcommand{\cdm}{{\textsc{cdm}}} %
\newcommand{\cwdm}{{\textsc{cwdm}}\xspace} %
\newcommand{\lcwdm}{\ensuremath{\Lambda}\textsc{CWDM}\xspace} %
\newcommand{\wdm}{\textsc{wdm}\xspace} %
\newcommand{\bary}{\textsc{b}\xspace} %
\newcommand{\tot}{\textsc{tot}\xspace} %
\newcommand{\m}{\textsc{m}\xspace} %
\newcommand{\rad}{\textsc{r}\xspace} %
\newcommand{\fwdm}{{\ensuremath{f_\wdm}}\xspace}%
\newcommand{\Fwdm}{{\ensuremath{F_\wdm}}\xspace}%
\newcommand{\fs}{\textsc{fs}} %
\newcommand{\fsh}{\textsc{fsh}} %
\newcommand{\eq}{\mathrm{eq}} %
\newcommand{\dw}{\textsc{nrp}} %
\newcommand{\lya}{Ly-$\alpha$\xspace} %
\newcommand{\anr}{\ensuremath{a_\mathrm{nr}}\xspace} %Time of non-relativ. transition
\newcommand{\aeq}{\ensuremath{a_\mathrm{eq}}\xspace} %Time of non-relativ. transition

\begin{document}

\title{
  \begin{flushright}
    \small \texttt{CERN-PH-TH/2008-234\\
      LAPTH-1290/08}
  \end{flushright}Lyman-$\alpha$ constraints on warm and on warm--plus--cold dark matter models}

\author{\normalfont Alexey~Boyarsky$^{a,b}$, Julien Lesgourgues$^{c,d,e}$,
  Oleg~Ruchayskiy$^c$,
  Matteo Viel$^{f,g}$\\
  \small\it $^{a}$ETHZ, Z\"urich, CH-8093,  Switzerland\\
  \small\it   $^{b}$Bogolyubov Institute for Theoretical Physics, Kiev 03680, Ukraine\\
  \small\it $^{c}$\'Ecole Polytechnique F\'ed\'erale de Lausanne,
  \small\it FSB/BSP/ITP/LPPC, CH-1015, Lausanne, Switzerland\\
  \small\it $^d$PH-TH, CERN, CH-1211 Geneve 23, Switzerland\\
  \small\it $^e$ LAPTH, Universit\'e de Savoie, CNRS,
  B.P.110, F-74941 Annecy-le-Vieux Cedex, France\\
  \small\it $^f$INAF -- Osservatorio Astronomico di Trieste, Via G.B.~Tiepolo
  11,
  I-34131 Trieste, Italy\\
  \small\it $^g$INFN -- National Institute for Nuclear Physics, Via Valerio 2,
  I-34127 Trieste, Italy\\}
% \footnotesize\textbf{File \jobname.tex. ${}$Revision: 1.27 ${}$}\\
\date{}%

\maketitle

\begin{abstract}
  We revisit Lyman-$\alpha$ bounds on the dark matter mass in $\Lambda$ Warm
  Dark Matter ($\Lambda$WDM) models, and derive new bounds in the case of
  mixed Cold plus Warm models ($\Lambda$CWDM), using a set up which is a good
  approximation for several theoretically well-motivated dark matter models.
  We combine WMAP5 results with two different Lyman-$\alpha$ data sets,
  including observations from the Sloan Digital Sky Survey. We pay a special
  attention to systematics, test various possible sources of error, and
  compare the results of different statistical approaches. Expressed in terms
  of the mass of a non-resonantly produced sterile neutrino, our bounds read
  $m_\dw \ge 8$~keV (frequentist 99.7\% confidence limit) or $m_\dw \ge
  12.1$~keV (Bayesian 95\% credible interval) in the pure $\Lambda$WDM limit.
  For the mixed model, we obtain limits on the mass as a function of the warm
  dark matter fraction $\Fwdm$. Within the mass range studied here
  ($5~\mathrm{keV} < m_\dw < \infty$), we find that any mass value is allowed
  when $\Fwdm < 0.6$ (frequentist 99.7\% confidence limit); similarly, the
  Bayesian joint probability on $(\Fwdm, 1/m_\dw)$ allows any value of the
  mass at the 95\% confidence level, provided that $\Fwdm < 0.35$.
\end{abstract}
\section{Introduction}
\label{sec:intro}

The nature of Dark Matter (DM) is one of the most intriguing questions
of particle astrophysics. Its resolution would have a profound impact
on the development of particle physics beyond its Standard Model (SM).
Although the possibility of having massive compact halo objects
(MACHOs) as a dominant form of DM is still under debates (see recent
discussion in~\cite{Calchi:07} and references therein), it is widely
believed that DM is made of non-baryonic particles.  Yet the SM of
elementary particles does not contain a viable DM particle candidate
-- a massive, neutral and stable (or at least cosmologically
long-lived) particle.  Active neutrinos, which are both neutral and
stable, form structures in a top-down
fashion~~\cite{Zeldovich:70,Bisnovatyi:80,Bond:80,Doroshkevich:81,Bond:83},
and thus cannot produce the observed large scale structure.
Therefore, the DM particle hypothesis implies an extension of the
SM. Thus, constraining the properties of DM helps to distinguish
between various DM candidates and may help to differentiate among
different models Beyond the Standard Model (BSM).

What is known about the properties of DM particles?

DM particle candidates may have very different masses (for reviews
of DM candidates see
e.g.~\cite{Bergstrom:00,Bertone:05,Carr:06,Taoso:07}): massive
gravitons with mass $\sim 10^{-19}\ev$~\cite{Dubovsky:04}, axions with
mass $\sim 10^{-6}\ev$~\cite{Wilczek:77,Weinberg:77,Holman:83},
sterile neutrinos with mass in the keV range~\cite{Dodelson:93},
supersymmetric (SUSY) particles (gravitinos~\cite{Pagels:82},
neutralinos~\cite{Lee:77,Haber:85}, axinos~\cite{Covi:99,Covi:01} with mass
ranging from a few eV to hundreds GeV), supersymmetric
Q-balls~\cite{Kusenko:97b}, WIMPZILLAs with mass $\sim
10^{13}\gev$~\cite{Kuzmin:98,Chung:99}, and many others. Thus, the
mass of DM particles becomes an important characteristics which may
help to distinguish between various DM candidates and, more
importantly, may help to differentiate among different BSMs.  A quite
robust and model-independent \emph{lower bound} on the mass of
spin-one-half DM particles was suggested in~\cite{Tremaine:79}. The
idea was based on the fact that for any fermionic DM the average
phase-space density (in a given DM-dominated, gravitationally bound
object) cannot exceed the phase-space density of the degenerate Fermi
gas.  This argument, applied to the most DM-dominated dwarf Spheroidal
(dSph) satellites of the Milky Way leads to the bound $m_\dm >
0.41\kev$~\cite{Boyarsky:08a}. For particular DM models (with known
primordial velocity dispersion) and under certain assumptions about
the evolution of the system during structure formation, this limit can
be strengthened.  This idea was developed in a number of
works~\cite{Madsen:84,Madsen:91,Madsen:90,Dalcanton:00,Hogan:00,Madsen:00,Madsen:01,Boyanovsky:08,Boyarsky:08a,Gorbunov:08b}
(for recent results and a critical discussion see~\cite{Boyarsky:08a}).

For any DM candidate there should exist a production mechanism in the
early Universe. Although it is possible that the DM is produced
entirely through interactions with non-SM particles (e.g. from the
inflaton decay) and is inert with respect to all SM interactions, many
popular DM candidates are produced via interactions with the SM
sector.  One possible class of DM candidates are \emph{weakly
  interacting massive particles} -- WIMPs~\cite{Lee:77}. They have
interactions with SM particles of roughly electroweak strength.  The
WIMPs are completely stable particles, as their interaction strength
would otherwise imply a lifetime incompatible with a sizable relic
density today. Therefore, they are produced in the early Universe via
annihilation into SM particles with roughly weak interaction
cross-section (see e.g.~\cite{Kolb:90}).  To have a
correct abundance, these particles should have a mass in the GeV to TeV
range. Such a mass and interaction strength means that they decouple
from the primeval plasma deeply in the radiation dominated epoch,
while being non-relativistic.  Such particles have a Boltzmann velocity
distribution function at the moment of freeze-out and are called
``cold DM'' (CDM). In the CDM scenario, structure formation goes in the
bottom-up fashion: the smallest structures (with supposed masses as small
as that of the Earth) form first, and then combined into larger
structures.

Another wide class of DM candidates can be generically called
\emph{superweakly} interacting (see e.g.~\cite{Feng:2003xh}), meaning that
their cross-section of interaction with SM particles is much smaller than the
weak one.  There are many examples of \emph{super-WIMP} DM models: sterile
neutrinos~\cite{Dodelson:93}, gravitinos in theories with broken
R-parity~\cite{Takayama:00,Buchmuller:07}, light volume
moduli~\cite{Conlon:07} or Majorons~\cite{Lattanzi:07}. Many of these
candidates (e.g. sterile neutrino, gravitino, Majoron) possess a two-body
decay channel: $\dm \to \gamma + \nu,\gamma+\gamma$.  These scenarios are very
interesting, since apart from laboratory detection such particles can be
searched in space. Looking for a monochromatic decay line in the spectra of
DM-dominated objects provides a way of \emph{indirect detection} of DM, and
could help to \emph{constrain} its interaction strength with SM particles. The
astrophysical search for \emph{decaying} DM is in fact more promising than for
stable (annihilating) DM. Indeed, the decay signal is proportional to the
\emph{column density} $\int\rho_\dm(r)dr$, i.e.  the density integrated along
the line of sight, and not the squared density $\int\rho^2_\dm(r)dr$ (as it is
the case for annihilating DM). As a consequence, \emph{(i)} a vast variety of
astrophysical objects of different nature would produce roughly the same decay
signal~\cite{Boyarsky:06c,Bertone:07}; \emph{(ii)} this gives more freedom for
choosing the observational targets, and avoid complicated astrophysical
backgrounds (e.g. one does not need to look at the Galactic center, expecting
a comparable signal from dark outskirts of galaxies, clusters and dark
dSph's); \emph{(iii)} if a candidate line is identified, its surface
brightness profile may be measured (since it does not decay quickly away from
the centers of objects), in order to distinguish it from astrophysical lines
(which usually decay in outskirts) and to compare with the signal from several
other objects. For all these reasons, astrophysical search for decaying DM can
almost be viewed as another type of \emph{direct detection}.  A search for DM
decay signals was conducted both in the keV -- MeV
range~\cite{Boyarsky:05,Boyarsky:06b,Boyarsky:06c,Boyarsky:06d,Boyarsky:06e,Boyarsky:06f}
and GeV range~\cite{Bertone:07}.

The small strength of interaction of light super-WIMP particles
usually implies that:
\begin{inparaenum}[\em (i)]
\item they were never in thermal equilibrium in the early Universe; and
  \item
they were produced deep in the radiation dominated epoch, while still being
relativistic (unlike CDM). 
\end{inparaenum}
The latter property means that the density perturbation of these
particles are suppressed for scales below their \emph{free-streaming
length},\footnote{The free-streaming length is the equivalent of the
Jeans length for a collisionless fluid. It is given by the ratio 
$\lambda_\fs \sim \langle v \rangle /H$, where $\langle v \rangle$
is the velocity dispersion of the particles.} which affects structure
formation on those scales. If the maximum free-streaming length
roughly coincides with galaxy scales, the particles are called Warm
Dark Matter (WDM), see e.g.~\cite{Bode:00}.

Currently, the best way to distinguish between WDM and CDM models is the
analysis of Lyman-$\alpha$ (Ly-$\alpha$) forest data~(for an introduction see
e.g.~\cite{Hui:97,Gnedin:01,Weinberg:03}).  This method requires particular
care. Part of the physics entering into the \lya analysis is not fully
understood (for a recent review see \cite{Meiksin:07}), and can be
significantly influenced by DM particles~(see
e.g.~\cite{Viel:2003fx,Kim:07a,Bolton:07a,Gao:07,Biermann:06,Stasielak:06,Faucher:07}).
The results can also be affected by approximations related to computational
difficulties.  Indeed, at the redshifts probed by Ly-$\alpha$ data, structures
are already undergoing non-linear evolution. In order to relate the measured
flux spectrum with the parameters of each cosmological model, one would have
to perform a prohibitively large number of numerical simulations.  Therefore,
various simplifying approximations have to be realized. Different approaches
are presented in~\cite{Viel:04,McDonald:05,Viel:05c}.

Previous works analyzing \lya constrains on
WDM~\cite{Hansen:01,Viel:05,Abazajian:05b,Seljak:06,Viel:06,Viel:07} assumed
the simplest $\Lambda$WDM model with a cut-off in the linear power spectrum
(PS) of matter density fluctuations. Such a PS arises when WDM particles are
\emph{thermal relics}.  However, in many super-weakly interacting DM models,
the PS has a complicated non-universal form due to the non-thermal primordial
velocity distribution~(see
e.g.~\cite{Feng:2004zu,Roszkowski:04,Cerdeno:2005eu,Choi:2008zq}).  For
example, in gravitino models with broken R-parity, gravitino production occurs
in two stages: thermally at high temperatures (see
e.g.~\cite{Bolz:00,Pradler:06,Rychkov:07}), and then non-thermally through the
late decay of the next-to-lightest supersymmetric particle~(see
e.g.~\cite{Borgani:96,Feng:2004zu}). Therefore the primordial velocity
distribution results from the superimposition of a colder (resonant) and a
warmer (thermal) component, and the PS is characterized by a step and a
plateau on small scales (rather than a sharp cut-off).

Another interesting super-WIMP WDM model with a non-trivial PS is the sterile
neutrino DM of the \numsm -- an extension of the SM by three sterile
(right-handed, gauge singlet) neutrinos~\cite{Asaka:05a,Asaka:05b}.  Having
masses of all new particles below the electro-weak scale, this model is able
to explain simultaneously three BSM phenomena:
\begin{inparaenum}[\em (i)]
\item the transition between the neutrino of different flavours
  (\emph{neutrino oscillations});
\item the production of a non-zero baryon number in the early Universe; and
\item the existence of DM candidate with correct relic
  density.\footnote{This model also eases the \emph{hierarchy
      problem}, as the \numsm can be thought of as a consistent
    quantum field theory, valid all the way to the Planck scale. The
    hierarchy problem is then shifted into the realm of quantum
    gravity, broadly understood as a fundamental theory, superseeding
    at the Planck scale~\cite{Shaposhnikov:07b}. }
\end{inparaenum}
Sterile neutrino DM in the \numsm is produced via its mixing
  with active neutrinos in the early
  Universe~\cite{Dodelson:93,Shi:98,Abazajian:01a,Asaka:06c,Laine:08a}. In the
  absence of lepton asymmetry, the production rate is strongly suppressed at
  temperatures above few hundreds of MeV and peaks roughly at $ T_{peak} \sim
  130\left(\frac{M_\dm}{1\kev}\right)^{1/3}$~\mbox{MeV}~\cite{Dodelson:93}.
  The resulting momentum distribution of sterile neutrino DM is approximately
  that of a rescaled Fermi-Dirac~\cite{Dodelson:93,Dolgov:00}, with its
  temperature equal to that of active neutrinos (see~(\ref{eq:fp}) below). We
  will call this production mechanism \textbf{NRP} -- \emph{non-resonant
    production}.  Notice that this production channel is always present
  (although it may be subdominant, see below).

  The sterile neutrino DM production changes in the presence of a lepton
  asymmetry~\cite{Shi:98}.  In this case the dispersion relations for active
  and sterile neutrinos may cross each other. This results in an effective
  transfer of an excess of active neutrinos (or antineutrinos) to the
  population of DM sterile neutrinos. The lepton asymmetry necessary for this
  \emph{resonant production} (\textbf{RP}) is generated via the decay of
  heavier sterile neutrinos~\cite{Shaposhnikov:08a}. In the RP scenario, for
  each mass and lepton asymmetry, the velocity dispersion has a colder
  (resonant) component and a warmer (non-resonantly produced) tail. For a
  recent review see e.g.~\cite{Boyarsky:09a}.

  Sterile neutrinos having colder and warmer velocity components can also be
  produced through the decay of a gauge-singlet scalar, followed by a
  subsequent non-resonant production~%
  \cite{Shaposhnikov:06,Kusenko:06a,Petraki:07,Petraki:08,Boyanovsky:08b} (see
  also \cite{Khalil:08,Wu:09} for other mechanisms of production of sterile
  neutrino).  The decay of this scalar field should occur at temperatures
  $\gtrsim 100$~GeV and the produced sterile neutrinos do not share subsequent
  entropy releases~\cite{Shaposhnikov:06,Petraki:07}. As a result the
  characteristic momentum of sterile neutrinos, produced in that way, is
  several times colder than the average momentum of the NRP component (see
  e.g.~\cite{Shaposhnikov:06}).
Notice that in this case the production is determined not only by the mixing
between active and sterile neutrinos, but also by the coupling of the sterile
neutrino with the scalar.

We see that in several well-motivated models, the primordial momentum
distribution of DM particles contains a warm (Fermi-Dirac-like component) and
a colder one. %
Therefore, in this work, apart from updating existing bounds on $\Lambda$WDM
models, we consider \lya bounds in the \emph{$\mathit{\Lambda}$CWDM} scenario,
including a mixture of a CDM and WDM. This model is characterized by two
independent parameters beyond the usual dark matter fraction $\Omega_{\rm
  DM}$: the velocity dispersion (or free-streaming length) of the WDM
component, and the fraction of WDM density. We will overview and provide a
critical analysis of uncertainties in the \lya method for this model, and
derive constraints on the above parameters.  Finally, we will discuss the
consequences of these results for some physically relevant DM models of this
kind.

The first qualitative analysis of \lya bounds in the $\Lambda$CWDM 
scenario was
performed in~\cite{Palazzo:07}. In this work, the authors
computed the \emph{linear} $\Lambda$CWDM power spectra
projected along the line of sight. The PS suppression (as
compared to the pure $\Lambda$CDM case) at some fiducial scale 
%($k=2\:h/\mathrm{Mpc}$)
was compared with that quoted in~\cite{Seljak:06} for several WDM
masses. A $\Lambda$CWDM model was
considered as ruled out at some confidence level if it produced the same
suppression in the 1D projected linear PS as a pure $\Lambda$WDM model
ruled out at the same level in~\cite{Seljak:06}.

This qualitative method should be improved in two directions. First of all,
the \lya method provides a measurement of the 1D \emph{non-linear} flux
spectrum.  The non-linear evolution ``mixes'' different scales. Therefore, two
linear PS passing through the same point at a given wave number but having a
different behavior otherwise would fit \lya data differently. Secondly, the
effects of CWDM suppression on the matter PS can be somewhat compensated by
the change of cosmological parameters (DM abundances, $\sigma_8$, etc.).
Therefore, a proper \lya analysis should include a simultaneous fitting of all
cosmological plus CWDM parameters to the \lya (and possibly other
cosmological) data. Such an analysis is performed in this work.

The paper is organized as follows.  
In section \ref{sec:plateau-cwdm-models}, we describe the effect of a
mixture of cold and warm dark matter on the linear matter power
spectrum, and present some analytic approximations.
In section \ref{sec:flux-power-spectrum}, we introduce the two \lya
data sets used in this analysis: the Viel, Haenelt \& Springel (VHS)
data compiled in Ref.~\cite{Viel:2004bf}, and the Sloan Digital Sky
Survey (SDSS) data complied in Ref.~\cite{McDonald:04b,McDonald:05}.
In section \ref{sec:veloc-init-cond}, we discuss the issue of
including thermal velocities in hydrodynamical simulations in presence
of warm dark matter.
The reader only interested in our results could go directly to section
\ref{sec:results-vhs}, where we present the bounds based on the
conservative VHS data set, and to section \ref{sec:sdss-results},
which contains our analysis for the more constraining SDSS data.
In section \ref{sec:syst-uncert}, we describe the systematic errors 
included in our analysis.
Finally, in section \ref{sec:conclusion}, we discuss some implications
of our results.

\section{Linear power spectrum in CWDM models}
\label{sec:plateau-cwdm-models}

The growth of matter density perturbations in the Universe is
conveniently described in terms of a \emph{power spectrum}
$P(k)=\langle | \frac{\delta \rho_\m}{\rho_\m}|_k^2\rangle$ (where
$\rho_\m$ is the matter density, and $<\cdot >$ denotes the
  ensemble average).  The time evolution of this power spectrum
  from its primordial form (taken in this paper to be a power law with
  index ${n_s}$) to a redshift of interest depends on cosmological
  parameters (fraction of dark $\Omega_\dm$ and baryonic
  $\Omega_\bary$ matter, $\Omega_\Lambda$, Hubble constant, etc.)

Pure warm DM models contain another parameter which affects the evolution of
density perturbations: the initial DM velocity dispersion. In this paper, for
definiteness, we assume that the WDM component has a mass $m_\dw$ and a
primordial phase-space distribution $f(p)$ of the form
\begin{equation}
  f(p) = \frac\chi{\exp(p/T_\nu)+1}~.
\label{eq:fp}
\end{equation}
This form is motivated by the scenario of non-resonantly produced
(NRP) sterile neutrinos~\cite{Dodelson:93,Dolgov:00}. The temperature
$T_\nu \equiv (4/11)^{1/3} T_\gamma$ is the temperature of active
neutrinos in the instantaneous decoupling approximation, while the
normalization parameter $\chi\ll 1$ is determined by the mass $m_\dw$
and the requirement of a correct DM
abundance\footnote{Eq.(\ref{eq:fp}) is only an approximation to the
real sterile neutrino DM spectrum, which can be computed only
numerically (see~\cite{Asaka:06c} for details).}. Another
well-motivated WDM model is that of \emph{thermal relics} (TR). This
is a generic name for particles that were in thermal equilibrium in
the early Universe and decoupled while being relativistic (unlike
sterile neutrinos which never equilibrated). In the TR case, the
phase-space distribution is of the Fermi-Dirac type: namely, similar
to equation~(\ref{eq:fp}) although $\chi$ is equal to one, and the
temperature $T_{\textsc{tr}}$ is not equal to $T_\nu$ but can be
deduced from the mass of thermal relics $m_\textsc{tr}$ and the DM
abundance.  At the level of linear perturbations, the NRP and TR
models are formally equivalent under the identification
$T_{\textsc{tr}} =\chi^{1/4} T_\nu$ and mass $m_{\textsc{tr}} =
\chi^{1/4} m_\dw$ \cite{Colombi:1995ze}.

To separate the influence of the velocity dispersion on the evolution
of density perturbations, it is convenient to construct a
\emph{transfer function} -- the square root of the ratio of the matter
power spectrum of a given $\Lambda$WDM model to that of a pure
$\Lambda$CDM with the same choice of cosmological parameters:
\begin{equation}
  \label{eq:7}
  T(k) \equiv  \left(\frac{P_{\Lambda\wdm}}{P_{\Lambda\cdm}}\right)^{1/2}~.
\end{equation}
This transfer function turns out to be rather insensitive to the values
of cosmological parameters, and is well fitted by
\begin{equation}
  \label{eq:30}
  T(k)
  = (1+(k/k_0)^{2\nu})^{-5/\nu}~,
\end{equation}
where $\nu=1.12$, and the scale $k_0$ (related to the free-steaming scale) is
given as a function of the mass and cosmological parameters (see
e.g.~\cite{Viel:05}). This function describes a cut-off with $k^{-10}$
dependence on scales $k \gg k_0$.

\begin{figure}[t]
  \centering
  \includegraphics[width=\textwidth]{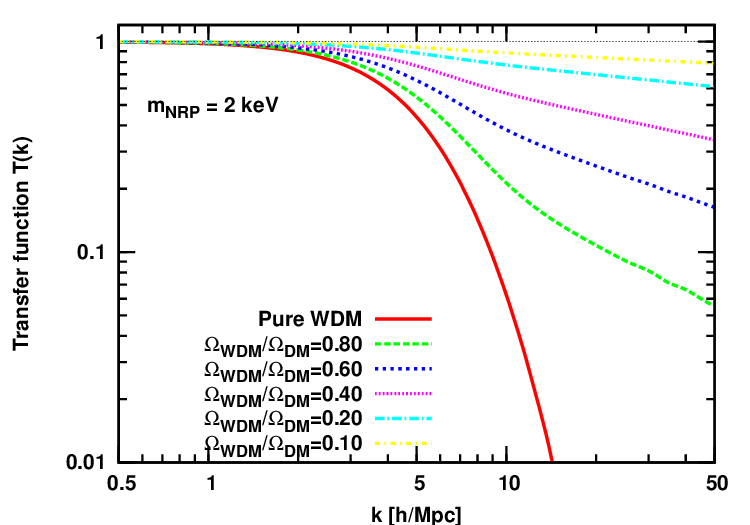} %
  \caption{Transfer functions 
$T(k) \equiv \left[{P_{\Lambda \cwdm}(k)} / {P_{\Lambda \cdm}(k)} \right]^{1/2}$
for $m_\dw = 2~\mathrm{keV}$ and different WDM
fractions: from left to right, $\Omega_\wdm / (\Omega_\wdm+ \Omega_\cdm) = 1,
0.8, 0.6, 0.4, 0.2, 0.1$. Other parameters are fixed to $\Omega_\bary=0.05$,
$\Omega_\m=0.3$, $\Omega_{\Lambda}=0.7$, $h=0.7$.}
  \label{fig:cwdm-m2000}
\end{figure}

\begin{figure}[t]
  \centering
  \includegraphics[width=\textwidth]{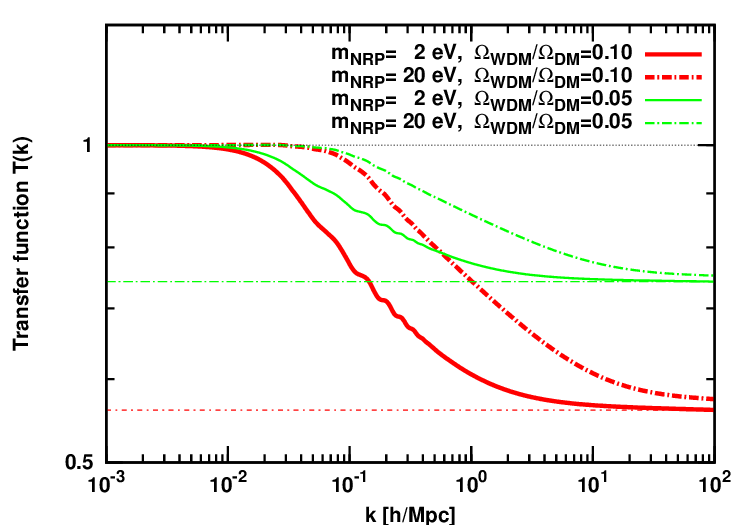}
  \caption{The height of plateau in the transfer function $T(k) \equiv
    \left[{P_{\Lambda \cwdm}(k)} / {P_{\Lambda \cdm}(k)} \right]^{1/2}$
    depends only on \fwdm only and does not change with the mass. In these
    examples, the mass is equal to $2 \, \ev$ (solid line) or $20 \, \ev$
    (dashed-dotted line), while $\Omega_\wdm / (\Omega_\wdm+ \Omega_\cdm) =
    0.1$ (red) or $0.05$ (green). Other parameters are fixed to
    $\Omega_\bary=0.05$, $\Omega_\m=0.3$, $\Omega_{\Lambda}=0.7$, $h=0.7$.}
  \label{fig:plateau_var_mass}
\end{figure}

In this paper we will consider a class of models where DM is a mixture of cold
and warm (with velocity dispersion in the form~(\ref{eq:fp})) components.  We
define the \emph{WDM fraction} \fwdm as
\begin{equation}
  \label{eq:3}
  \fwdm \equiv \frac{\Omega_\wdm}{\Omega_\m}, \qquad \Omega_\m = \Omega_\cdm + \Omega_\wdm + \Omega_\bary
\end{equation}
(note the presence of the baryonic fraction $\Omega_\bary$ in the
denominator).  The transfer function of $\Lambda$CWDM models is
qualitatively different.  The characteristic behavior of the
$\Lambda$CWDM power spectrum as compared to the pure $\Lambda$CDM one
for the same cosmological parameters is shown on
Figs.~\ref{fig:cwdm-m2000}, \ref{fig:plateau_var_mass}, and can be
summarized as follows:
\begin{compactitem}[--]
\item The scale where the transfer function departs from 1 depends on the mass
  $m_\dw$ of the WDM component, but not on $\fwdm$, as
  Figs.~\ref{fig:cwdm-m2000} and \ref{fig:plateau_var_mass} demonstrate.
\item For large $k$'s, the transfer function approaches a constant
  plateau. The height of this plateau is determined solely by \fwdm,
  as Fig.~\ref{fig:plateau_var_mass} demonstrates.
\end{compactitem}

\subsection{Characteristic scales}
\label{sec:char-scal}

At any given time, the comoving free-streaming wavenumber of the warm
component is defined as
\begin{equation}
k_\fs \equiv \sqrt{4 \pi G \rho} \, \frac{a}{\langle v \rangle}
= \sqrt{\frac{3}{2}} \, \frac{a H}{\langle v \rangle}
\end{equation}
where $\rho$ is the total density of the universe at the time
considered, $\langle v \rangle$ is the velocity dispersion of the warm
component only (not to be confused with the velocity dispersion of the
total cold plus warm component, which depends on $\fwdm$), $a$ is the
scale factor and $H$ the Hubble rate.  This definition is analogous to
that of the Jeans scale for a fluid. The wavenumber $k_\fs$ appears in
the perturbation equations, and corresponds to the wavelength below
which perturbations cannot experience gravitational collapse due to
their velocity dispersion.  In the radiation era, when the fluid is
ultra-relativistic, $\langle v \rangle \simeq 1$ (using units where
$c=1$) and $k_\fs$ evolves like the comoving Hubble scale, $k_\fs =
\sqrt{3/2} \,a H \propto t^{-1/2}$.  After the non-relativistic
transition, $\langle v \rangle = \langle p \rangle / m$ scales like
$a^{-1}$; with our choice for $f(p)$, the average velocity is given
approximately by $\langle v \rangle = 3 T_{\nu} / m_\dw$. Hence,
during the rest of radiation domination, the comoving free-streaming
scale remains constant with
\begin{equation}
k_\fs = \sqrt{4 \pi G \rho_R} \, \frac{a m_\dw}{3 T_{\nu}}
= \sqrt{ \frac{\Omega_R}{6} } H_0 \frac{m_\dw}{T_{\nu}^0}
= 7 \left( \frac{m_\dw}{1~\mathrm{keV}} \right) \left( \frac{0.7}{h} \right) 
h/\mathrm{Mpc}
\label{kfs_const}
\end{equation}
where $T_{\nu}^0$ is the current neutrino temperature, and we assumed
the scale factor to be normalized to unity today. Later on, $k_\fs$
increases like $t^{1/3}$ during matter domination, and even faster
during $\Lambda$ domination. Today, assuming the universe to be flat,
the free-streaming scale is equal to
\begin{equation}
k_\fs^0 = \sqrt{4 \pi G \rho_c^0} \, \frac{m_\dw}{3 T_{\nu}^0}
= \sqrt{ \frac{1}{6} } H_0 \frac{m_\dw}{T\, _{\nu}^0}
= 8 \times 10^2 \left( \frac{m_\dw}{1~\mathrm{keV}} \right) \,
h/\mathrm{Mpc}~.
\end{equation}
Below this scale (for $k > k_\fs$), perturbations can never experience
gravitational clustering. Hence, this scale marks roughly the
beginning of the region for which the free-streaming effect is maximal
and the CWDM power spectrum is reduced by a constant amplitude (the
plateau in Fig.~\ref{fig:plateau_var_mass}).

By analogy with the case of light active neutrinos, one could expect that the
minimum of the free-streaming scale $k_\fs^{\rm min}$ (Eq.~(\ref{kfs_const}))
would give the point at which the $\Lambda$CWDM power spectrum starts to
differ from the $\Lambda$CDM one. However, $k_\fs^{\rm min}$ does not give the
right answer, as can be checked on figure Fig.~\ref{fig:plateau_var_mass}. In
fact, the largest scale affected by free-streaming is nothing but the present
value of the particle horizon of warm particles with a typical velocity
$\langle v \rangle$, that we will call the (comoving) free-streaming horizon
$\lambda_\fsh$
\begin{equation}
\lambda_\fsh^0 
= \int_0^{t_0} \frac{\langle v \rangle}{a} ~dt
= \int_0^1 \frac{\langle v \rangle}{a^2 H} ~da~.
\end{equation}
This scale can be computed easily if we neglect the impact of
$\Lambda$ (it is easy to check that the contribution of the
$\Lambda$-dominated stage to the above integral is completely negligible;
actually, neglecting also the matter-dominated stage would be
sufficient for an order of magnitude estimate).  We can decompose the
integral as:
\begin{equation}
\lambda_\fsh^0 =
\int_0^{\anr} \frac{da}{a^2 H} 
+ \frac{3 T_\nu^0}{m_\dw} \int_{\anr}^1 \frac{da}{a^3 H}
\end{equation}
with $\anr = T_\nu^0/ m_\dw$ being the scale factor at the time of the
non-relativistic transition. Using the Friedman equation, this becomes
\begin{equation}
\lambda_\fsh =
\frac{1}{\sqrt{\Omega_R} H_0} \int_0^{\anr} da
+ \frac{3 \anr}{\sqrt{\Omega_R} H_0} \int_{\anr}^1 \frac{da}{a \sqrt{1+\aeq/a} }
\end{equation}
where $\aeq = \Omega_R/\Omega_M$ is the scale factor at the time of
matter/radiation equality. In the limit $\anr \ll 1$, the result is simply
\begin{equation}
\lambda_\fsh^0 = \frac{\anr}{\sqrt{\Omega_R} H_0} \left( 1 + 
6 \, \mathrm{Arcsinh} \sqrt{\frac{\aeq}{\anr}} \right) \simeq
\frac{\anr}{\sqrt{\Omega_R} H_0} \left( 1 + 
3 \ln \left[ 4 \frac{\aeq}{\anr}\right] \right)
\end{equation}
which corresponds to the wavenumber
\begin{equation}
k_\fsh^0 = \sqrt{\Omega_R} H_0 \frac{m_\dw}{T_\nu^0} \left( 1 + 
3 \ln \left[ 4 \frac{\Omega_R}{\Omega_M} \frac{m_\dw}{T_\nu^0} \right] \right)^{-1}
\end{equation}
Taking $\Omega_R h^2 =4 \times 10^{-5}$ we finally obtain
\begin{equation}
  k_\fsh^0 = 0.5 
  \left(\frac{m_\dw}{1~\mathrm{keV}}\right) 
  \left(\frac{0.7}{h}\right)
  \left( 1 + 0.085 \ln \left[
      \left(\frac{0.1}{\Omega_M h^2}\right) \left(\frac{m_\dw}{1~\mathrm{keV}}\right) \right]
  \right)^{-1} h/\mathrm{Mpc}~.
\end{equation}
One can check in Fig.~\ref{fig:plateau_var_mass} that this corresponds
indeed to the scale at which the $\Lambda$CWDM power spectrum starts
to differ from the $\Lambda$CDM one. Note that the distinction between
the free-streaming horizon and the free-streaming scale is not
commonly used, because in the case of active neutrinos (which become
non-relativistic during matter domination) one has $k_\fsh^0 \sim
k_\fs^{\rm min}$; in this case, it is not necessary to introduce the
quantity $k_\fsh$, and the scale marking the beginning of the step is
usually said to be $k_\fs^{\rm min} = k_\fs^\mathrm{nr}$. In the case
of WDM or CWDM, the distinction is important, because between
$t_\mathrm{nr}$ and $t_{\rm eq}$ the free-streaming scale $\lambda_\fs$ remains 
constant, while the free-streaming horizon
$\lambda_\fsh$ increases typically  by one order of magnitude.

\subsection{Plateau in CWDM}
\label{sec:plateau-cwdm}

We can try to estimate the amplitude $T_\mathrm{plateau}$ of the small-scale
power spectrum suppression,
\begin{equation}
T_\mathrm{plateau} \equiv \left[ \frac{P_{\Lambda \cwdm}(k)}{P_{\Lambda \cdm}(k)} \right]^{1/2}
\qquad
{\rm for}
\qquad
k \gg k_\fs^0 =  8 \times 10^2 \left( \frac{m_\dw}{1~\mathrm{keV}} \right) \,
h/\mathrm{Mpc}~.
\label{eq:sdef}
\end{equation}
Let us consider a $\Lambda$CDM and a $\Lambda$CWDM model sharing the
same cosmological parameters, and in particular the same $\Omega_\dm$ and
$\Omega_\bary$.
For simplicity, we neglect the mass of active neutrinos throughout
this work. Until the time at which the warm component of the
$\Lambda$CWDM model becomes non-relativistic, these models are
indistinguishable.  After this transition, but still before
matter/radiation equality, the evolution of the gravitational
potential is fixed by that of the photon and neutrino
components, which dominate over dark matter; hence, the (slow) growth
of cold dark matter inhomogeneities, $\delta_\cdm \equiv \delta
\rho_\cdm / \rho_\cdm$, is independent of $\fwdm$. We
conclude that for any wavenumber $k$, $\delta_\cdm(a_\eq)$ is the same
in the two cases $\fwdm=0$ and $\fwdm \neq 0$.

After equality, the evolution of the gravitational potential is fixed
by the clustering of the dark matter and baryon components. Inside
the Hubble radius, the growth of $\delta_\cdm$ can be inferred from the
Newtonian equation of evolution
\begin{equation}
\ddot{\delta}_\cdm + \frac{\dot{a}}{a} \dot{\delta}_\cdm
= 4 \pi {\cal G} a^2 \delta \rho_\tot
= \frac{3}{2} \left( \frac{\dot{a}}{a} \right)^2 
\frac{(\rho_\cdm + \rho_\bary) \delta_\cdm
+ \rho_\wdm \delta_\wdm}{\rho_\tot}
\label{eq:newton}
\end{equation}
where we used the fact that $\delta_\cdm = \delta_\bary$ soon after 
equality, and a dot denotes a derivative with respect to conformal time.
In the $\Lambda$CDM model (with $\rho_\wdm=0$), this equation has a well-known
solution leading today to
\begin{equation}
  \label{eq:grcdm}
  \delta_\cdm(a_0) = \left(\frac{a_0 g(a_0)}{a_{\eq}}\right)\delta_\cdm(a_{\eq})
\end{equation}
where the constant $g(a_0) \leq 1$ is determined by the value of
$\Omega_\Lambda = 1 - \Omega_\m$, and describes the reduction
in the perturbation growth rate
during the $\Lambda$ dominated epoch. It is roughly approximated by 
\begin{equation}
  g(a_0) \simeq \Omega_\m^{0.2}\left(1+0.003(\Omega_\Lambda/\Omega_\m)^{4/3}\right)~.
\end{equation}
(see e.g.~\cite{Lesgourgues:06} for details). A more precise numerical evaluation gives $g(a_0) = 
0.779$ for $\Omega_\Lambda = 0.7$.
The product $a \, g(a)$
is usually called the linear growth factor.

In the $\Lambda$CWDM model,
$\rho_\wdm$ cannot be neglected in the term $\rho_\tot$ appearing
in equation (\ref{eq:newton}),
but the warm component never clusters on small scales 
with $k > k_\fs^0$,
so that $\delta_\wdm$ can be set to zero. Then, a factor 
\begin{equation}
\frac{\rho_\cdm + \rho_\bary}{\rho_\tot}
\end{equation}
appears in the right hand-side of equation (\ref{eq:newton}). During matter
domination, this factor is equal to $(1 - \fwdm)$
and the linear growth factor can be found exactly (like in the case
of $\Lambda$CHDM models with massive active neutrinos):
\begin{equation}
  \delta_\cdm(a) = 
\left(\frac{a}{a_{\eq}}\right)^{1-(3/5) \fwdm}
\delta_\cdm(a_{\eq})~.
\end{equation}
During $\Lambda$ domination, there is no simple analytic result. In the case
of $\Lambda$CHDM, it has been shown in Ref.~\cite{Lesgourgues:06}
that a very good approximation is found
by just inserting the function $g(a)$ inside the parenthesis. In our case,
the same ansatz would give
\begin{equation}
  \delta_\cdm(a_0) = 
\left(\frac{a_0 g(a_0)}{a_{\eq}}\right)^{1-(3/5) \fwdm}
\delta_\cdm(a_{\eq})
\end{equation}
leading to a reduction in clustering with respect to the $\Lambda$CDM model
equal to
\begin{equation}
\frac{\delta_\cdm^\fwdm(a_0)}{\delta_\cdm^{\fwdm=0}(a_0)}
= \left(\frac{a_0 g(a_0)}{a_{\eq}}\right)^{-(3/5) \fwdm}~.
\label{eq:reduc}
\end{equation}
The power spectrum $P(k)$ is defined as the variance of the total
matter perturbation
\begin{equation}
\frac{\delta \rho}{\rho} = \frac{\rho_\bary \delta_\bary + \rho_\cdm
\delta_\cdm + \rho_\wdm \delta_\wdm}{\rho_\bary + \rho_\cdm + \rho_\wdm}~.
\end{equation}
Today, $\delta_\bary = \delta_\cdm$, and in the $\Lambda$CWDM model
$\delta_\wdm \ll \delta_\cdm$ for $k > k_\fs^0$. So, on such small scales, on has
\begin{equation}
P_{\Lambda \cdm}(k) = \langle \delta_\cdm^2 \rangle~,
\qquad
P_{\Lambda \cwdm}(k) = (1-\fwdm)^2 \langle \delta_\cdm^2 \rangle~.
\label{eq:spectra}
\end{equation}
Using equations (\ref{eq:reduc}) and (\ref{eq:spectra}),
the
suppression factor defined in eq.~(\ref{eq:sdef}) is found to be:
\begin{equation}
T_\mathrm{plateau} = (1 - \fwdm) \left(\frac{a_0 g(a_0)}{a_{\eq}}\right)^{-(3/5) \fwdm}~.
\label{eq:sresult}
\end{equation}
This equation is slightly different from the equivalent result in
$\Lambda$CWDM models, due to the fact that the WDM component becomes
non-relativistic during radiation domination.

\begin{figure}[t]
  \centering
  \includegraphics[width=\textwidth]{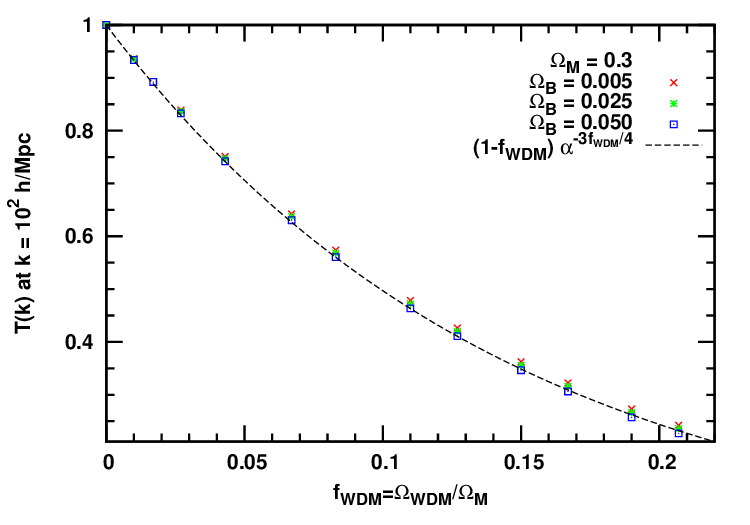}
  \caption{Suppression factor $T_\mathrm{plateau}=[{P_{\Lambda
        \cwdm}(k)}/{P_{\Lambda \cdm}(k)}]^{1/2}$ computed today at $k
        = 10^2 \: h/\mathrm{Mpc}$ with the Boltzmann code
        \textsc{CAMB}, for different values of $\fwdm \equiv
        \Omega_\wdm / \Omega_\m$ and $\Omega_\bary$. Other parameters are kept fixed: $\Omega_\Lambda=0.7$,
        $\Omega_\m=0.3$, $h=0.7$.  The dashed curve corresponds to the
        analytic prediction of Eq.~(\ref{eq:tfit}) with $\alpha \equiv
        {a_0 g(a_0)}/{a_{\eq}} = 2.75 \times 10^3$.}
  \label{fig:plato}
\end{figure}

Equation (\ref{eq:sresult}) encodes the right dependence of $T_\mathrm{plateau}$ on
$\fwdm$, but cannot fit accurately the numerical results of a
Boltzmann code for two reasons:

\begin{itemize}[--]
\item
first, in the above solution, we treated the transition from radiation
to matter domination as an instantaneous process. In fact, matter
comes to dominate in a progressive way, and $\delta_\cdm^\fwdm(a)$
starts to depart from $\delta_\cdm^{\fwdm=0}(a)$ slightly {\it before}
the time of equality.  Hence, the ratio $\delta_\cdm^\fwdm(a_\eq) /
\delta_\cdm^{\fwdm=0}(a_\eq)$ is actually smaller than one and
decreases when $f_\wdm$ increases. This means that equation
(\ref{eq:sresult}) slightly underestimates the effect of WDM.  Empirically,
we found that this can be corrected by increasing the exponential factor
from $3/5$ to $3/4$ (at least, as long as $\Omega_\m$
does not depart too much from 0.3).  In figure \ref{fig:plato}, we
compare the result of a full Boltzmann code calculation (using
\textsc{CAMB} \cite{Lewis:99}) with the empirical fit
\begin{equation}
T_\mathrm{plateau} = (1 - \fwdm) \left(\frac{a_0 g(a_0)}{a_{\eq}}\right)^{-(3/4) \fwdm}
\label{eq:tfit}
\end{equation}
and find very good agreement. In this example, 
\begin{equation}
\alpha \equiv \frac{a_0 g(a_0)}{a_{\eq}} =
\frac{\Omega_\m}{\Omega_\rad} g(a_0) = \frac{\Omega_\m h^2
g(a_0)}{4.16 \times 10^{-5}} = 2.75 \times 10^3
\end{equation}
for $T_\textsc{cmb} = 2.726$~K, $\Omega_\Lambda=0.7$, 
$\Omega_\m=0.3$, $h=0.7$.
\item second, we assumed that in all cases one has $\delta_\bary
\simeq \delta_\cdm$ soon after equality. However, we are interested
here in very small scales ($k \sim 100 h / \mathrm{Mpc}$) for which
the baryon and CDM overdensities do not converge towards each other
before the end of matter domination. Hence, the approximations
$\delta_\cdm \propto a$ or $\delta_\cdm \propto a^{1-(3/5) \fwdm}$
only apply to a short period: at the beginning of matter domination,
baryons do not contribute to the right-hand side of equation
(\ref{eq:newton}), and the growth of $\delta_\cdm$ is slower.  Since
this effect (governed by the parameter $\Omega_\bary / \Omega_\m$) is
similar in the two cases $\fwdm \neq 0$ and $\fwdm=0$, one does not
expect a strong dependence of $T_\mathrm{plateau}$ on the baryon
abundance. As can be seen in figure \ref{fig:plato}, the Boltzmann
code calculation shows that even radical variations in $\Omega_\bary$
hardly affect the behavior of the suppression factor. We conclude that
equation (\ref{eq:tfit}) is accurate up to a few percent for realistic
$\Lambda$CWDM models\footnote{This statement is meant for the plateau scales
$k \sim 100 h / \mathrm{Mpc}$. At much larger $k$, the situation becomes more complicated,
since according to the linear theory one has $\delta_\bary < \delta_\cdm$ even today.}.
\end{itemize}

For small \fwdm, one
can expand equation~(\ref{eq:tfit}) to obtain:
\begin{equation}
  \label{eq:11}
  T_\mathrm{plateau}^2 \approx 1-2\fwdm -\frac64 \fwdm \log \alpha \approx 1-14\fwdm~.
\end{equation}
This expression can be compared with $T_\mathrm{plateau}^2 \approx 1-8f_\nu$ obtained in the case of
active neutrinos~\cite{Hu:1997mj}.

\begin{figure}[t]
  \centering
  \includegraphics[width=0.49\textwidth]{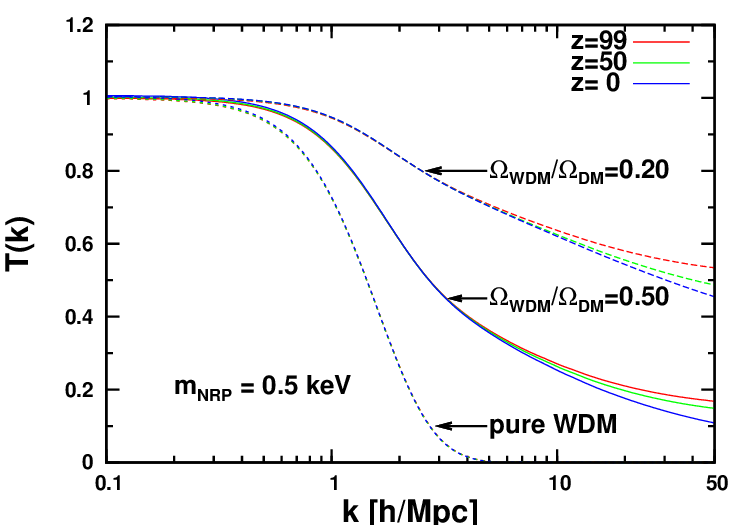}
  \includegraphics[width=0.49\textwidth]{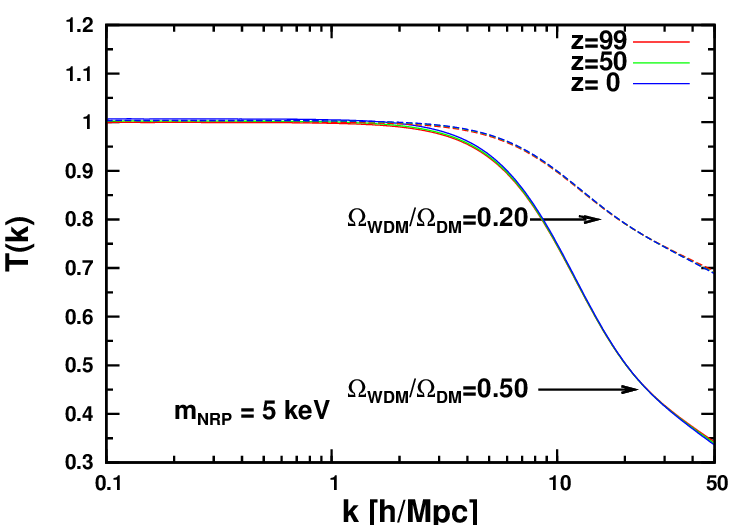}
  \caption{Variation of $T(k)=[{P_{\Lambda \cwdm}(k)}/{P_{\Lambda
        \cdm}(k)}]^{1/2}$ with redshift, for various values of $m$ and
        $\Omega_\wdm / (\Omega_\wdm+ \Omega_\cdm)$; other parameters
        are fixed to the same values as in previous figures. {\it
        (Left)} For $m=0.5~\mathrm{keV}$, the variation is clearly
        visible at large $k > k_\fs(z)$. {\it (Right)} For
        $m=5~\mathrm{keV}$, the free-streaming wavenumber is too large for
        any variation to be seen on the scales displayed here. }
  \label{fig:tkz}
\end{figure}

Note that equation (\ref{eq:tfit}) applies today, but the suppression
factor at higher redshift can simply be obtain by replacing $a_0$ by
$a=a_0/(1+z)$; so, it is clear that the transfer function $T(k)$ is
redshift-dependent, with a smaller step-like suppression at large $z$.
The function $T(k)$ is only expected to vary with $z$ on scales such
that $\delta_\wdm(z) < \delta_\cdm(z)$, i.e. for $k > k_\fs(z)$. This redshift dependence is illustrated in figure \ref{fig:tkz}.

\section{\lya method}
\label{sec:flux-power-spectrum}

In order to constrain cosmological models, it is always preferable to
rely on data dealing with linear scales. However, we have seen in the
previous section that the difference between $\Lambda$CDM and
$\Lambda$CWDM (or $\Lambda$WDM) models appears only on small scales
(typically, $k > 0.1$~h/Mpc) which cannot be probed by CMB
experiments, or by the reconstruction of the galaxy power spectrum in
the range in which the light-to-mass bias can be treated as
scale-independent. Hence, at the moment, the only way to discriminate
between $\Lambda$CDM and $\Lambda$CWDM (or $\Lambda$WDM) models is to
use observations of Lyman-$\alpha$ (\lya) absorption in the spectrum
of distant quasars (QSOs), which can be used as a tracer of
cosmological fluctuations on scales $k \sim (0.1-10) \, h \mpc^{-1}$,
at redshifts $z \sim (2-4)$
where the cosmological perturbations are already affected
by the non-linear evolution -- although not as much as today.

The \lya forest is a dense set of absorption lines observed in QSO spectra. It
is well established by analytical calculations and hydrodynamical simulations
that these absorption lines correspond to \lya absorption ($1s\to 2p$) by
clouds of neutral hydrogen at different redshifts along the line of sight.
This neutral hydrogen is part of the warm ($\sim 10^4$~K) and photoionized
intergalactic medium (IGM). Opacity fluctuations in the spectra arise from
fluctuations in the neutral hydrogen density, from which it is possible to
infer fluctuations in the total matter
distribution~\cite{bi,Viel:2001hd,Zaldarriaga:2001xs}. For each quasar, the
observed spectrum $I(z)$ can be expanded in (one-dimensional) Fourier space.
The expectation value of the squared Fourier spectrum is called the \emph{flux
  power spectrum}.  The data provide an estimate of the flux power spectrum
$P_F$ for various redshifts $z \sim 2-4$ (for which light rays of wavelength
$\lambda_{obs} = (1+z)1265\AA$ pass through the optical window of the Earth's
atmosphere). At these redshifts, the density perturbations of scales $k \sim
(0.1-10) \, h \mpc^{-1}$ already entered into a mildly non-linear stage of
gravitational collapse ($\delta \rho/ \rho\gtrsim1$).

The linear matter power spectrum $P(k)$ and the flux power spectrum $P_F(k)$
are complicated functions of the cosmological parameters.  In the rather
narrow range of scales probed by \lya data, the effects of an admixture of WDM
might be compensated by a change in other cosmological parameters ($\sigma_8$,
$\Omega_\m h^2$, $n_s$, etc.), or even in extra astrophysical parameters in
the case of the flux power spectrum. Therefore, it is important to perform
collective fits to the \lya data and other data sets (e.g. CMB). Cosmological
parameter extraction from combined cosmological data sets can be conveniently
performed with a Monte-Carlo Markov Chain (MCMC) technique, using e.g. the
public code {\sc CosmoMC}~\cite{Lewis:02}).

In order to predict $P_F(k,z)$ for a given cosmological model, it is
necessary to perform hydrodynamical simulations (while for CMB
experiments probing the linear matter power spectrum, it is sufficient
to compute the evolution of linear cosmological perturbations using a
Boltzmann code like {\sc camb}).
Hydrodynamical simulations are necessary both for simulating the
non-linear stage of structure formation and for computing the
evolution of thermodynamical quantities (in order to relate the
non-linear matter power spectrum to the observable flux power
spectrum).  The characteristic number of samples required for a
reliable determination of parameter probabilities ranges from $10^3$
for the simplest cosmological scenario to $10^4$ (or even $10^5$) when
more parameters and freedom are introduced in the model.  In
principle, in order to fit Lyman-$\alpha$ data, one should perform a
full hydrodynamical simulation for each Monte-Carlo sample. This is
computationally prohibitive, and various simplifying approximations
have been
proposed~\cite{Theuns:98,Gnedin:01,McDonald:05,Viel:04,Viel:05,Viel:05b,Viel:05c,Regan:06a}. Below,
we discuss the two datasets used in this paper, for which different
strategies have been implemented.

\subsection{VHS data}
\label{sec:vhs-data}

The VHS dataset was compiled in Ref.~\cite{Viel:2004bf} using 57 QSO spectra
from LUQAS \cite{Kim:2003qt} at $z\sim2.1$, and from Ref.~\cite{Croft:2000hs}
at $z\sim2.7$. Each spectrum was observed with high resolution and high
signal-to-noise, resulting in clean and robust measurements. However, this
dataset is based on a relatively small number of spectra, leading to a large
statistical uncertainty. Systematic errors were estimated in a conservative
way, and the flux power spectrum was only used in the range
$0.003~\mathrm{s/km} < k < 0.03~\mathrm{s/km}$ (roughly corresponding to
scales $0.3~\mathrm{h/Mpc} < k < 3~\mathrm{h/Mpc})$: at larger scales, the
errors due to uncertainties in fitting a continuum (i.e. in removing the long
wavelength dependence of the spectrum emitted by each QSO) become very large,
while at smaller scales, the contribution of metal absorption systems becomes
dominant.  It was shown in Ref.~\cite{Viel:2004bf} that the dependence of the
bias function $b(k,z) \equiv P_F(k,z) / P(k,z)$ on cosmological parameters can
be neglected for this data set, given the large systematic plus statistical
errorbars in VHS, and the limited range in $k$-space and $z$-space probed by
this data~\cite{Gnedin:01}.  Hence, this bias function can be computed for a
fiducial model, and the data points $P_F(k_i,z_j)$ can be translated in linear
spectrum measurements $P(k_i,z_j)$.  This approach is implemented in the
public module {\tt lya.f90} \cite{Viel:05} distributed with {\sc CosmoMC}.  We
will use this dataset in section \ref{sec:results-vhs} to derive conservative
estimates.

\subsection{SDSS data}
\label{sec:data:sdss}

For the Ly-$\alpha$ data obtained by the SDSS collaboration
\cite{McDonald:04b}, a different approach has been implemented in
Ref.~\cite{McDonald:05}.  The data points were obtained from $3035$
quasar spectra with low resolution and low signal-to-noise,
spanning a wide range of redshifts ($z=2.2-4.2$). Dealing
with low resolution and low signal-to-noise QSO spectra and extracting the flux power is a very difficult task
as shown in ~\cite{McDonald:05}. We refer to their analysis for a
comprehensive study of continuum fluctuations removal, metal lines
contamination, presence of damped \lya systems, resolution of the
spectrograph and noise level in each redshift bins. All these effects
need to be properly taken into account, and not modelling them
properly would impact the obtained flux power in a way that is
difficult to predict. In the following, we will use the flux power
provided by the SDSS collaboration, introducing nuisance parameters
for the resolution and noise in each redshift bin as suggest by
\cite{McDonald:04b,McDonald:05}, and implicitly assuming that all the
contaminants above have been either removed or properly modelled in
their analysis.

The low resolution of SDSS data implies that the flux power spectrum cannot be
measured on small scales (with $k \geq 0.02$~s/km). However, on larger scales,
the low resolution and signal-to-noise are compensated by
the statistics: with such a large number of quasars, statistical
errorbars on the flux power spectrum are much smaller than for VHS
data. Hence, using a fixed bias function would be
inappropriate. Instead, Ref.~\cite{McDonald:04b,McDonald:05} used a
large number of fast hydro-particle-mesh (HPM) simulations
\cite{Gnedin:1997td,Viel:05b} densely sampling the cosmological
parameter space, calibrated from a small number of full hydrodynamical
simulations.  These HPM runs could produce an inaccurate flux spectrum
estimation for cosmological models departing from the fiducial points
where the hydrodynamical simulations were performed, and a more
reliable calibration is probably needed.

A different approximation scheme was used in Ref.~\cite{Viel:05c}
using the same data. For each wavenumber and redshift, the flux power
$P_F(k,z)$ was Taylor-expanded around a
fiducial model at first order in cosmological parameters, using of the
order of $N$ full hydrodynamical simulations (here $N \sim 20$ is the
number of relevant cosmological/astrophysical parameters).  This
method can also become inaccurate far from the fiducial model,
especially when the likelihood and/or parameter posterior probability
deviates significantly from a multivariate Gaussian
distribution. However, in both methods, it can be checked a posteriori
that the inaccuracy introduced by interpolating/extrapolating around
the points where hydrodynamical simulations were performed is
negligible with respect to data error bars, at least in the range of
cosmological models producing reasonable fits to the data.

We deal with uncertainties related to \lya physics in the same way as in
Ref.~\cite{McDonald:05,Viel:05c}. Namely, we use 9 ``astrophysical'' nuisance
parameters.  Two describe the effective evolution of the optical depth, six
describe the evolution of the parameters $\gamma$ and $T_0$ entering the
relation between the temperature of the intergalactic medium and its
overdensity, $T=T_{0} (1+\delta)^{\gamma-1}$.  The parameter $T_0$ was allowed
to vary in the range $0 \le T_0 \le 10^5$~K, while $0\le \gamma\le 2$. Note
that the range of parameters chosen is very wide and embraces also the recent
results of \cite{Bolton:07a} that found evidences of an inverted ($\gamma <
1$) temperature-density relation. The parameters describing the thermal
evolution are modelled as a broken power-law at $z=3$ as
$y=A\,(\frac{1+z}4)^S$, with one parameter for the amplitude and two for the
slopes at $z<3$ and $z>3$ (the slopes are left free to vary in the range
$[-6,6]$). This is done in order to better mimic more complex evolutions able
for example to reproduce the Helium II reionization at $z\sim 3$ as suggested
by SDSS observations \cite{Bernardi:03}. The evolution of the effective
optical depth is modeled as a single power-law (two parameters: one for the
amplitude and one for the slope) within a very wide observational range.
Another parameter describes the contribution of strong absorption systems
(Damped Lyman-$\alpha$ systems) and this is left free following exactly the
suggestions of Ref.~\cite{McDonald:05} and applying the correction of
Ref.~\cite{McDonald:05syst}.

Finally, we have a total of 9 additional parameters that model
  uncertainties in the correction to the data -- eight parameters for the
  noise correction in various redshift bins, and one parameter describing the
  error of the resolution correction (these parameters do have strong a-posteriori priors as in Ref.~\cite{McDonald:05}).  All these parameters as well as those
  describing corrections for damped/high column density systems are
  constrained as suggested in~\cite{McDonald:05syst} and described
  in~\cite{Viel:05c}.

When deriving bounds on the parameters of interest
  below, we marginalized over these 18 parameters, and the final
  estimates of the astrophysical parameters are within the
  observational bounds.

  In order to take into account the contribution of SiIII metal lines to the
  power spectrum, we followed an approach similar to the one suggested in
  \cite{McDonald:05}. Namely, we introduced an extra parameter, describing the
  contribution of SiIII lines to the power spectrum, and fixed it to the
  maximum correction suggested by the SDSS collaboration~\cite{McDonald:05},
  motivated by the following arguments: i) the metal lines contribution peaks
  at smaller scales than those investigated here; ii) at large scales the
  effect is an overall suppression of power which is already taken into
  account by the fact that we are considering a wide range of possible mean
  effective optical depths (\cite{Kim:2003qt}); iii) the effect of modelling
  further the SiIII metal contribution by allowing $z$-dependence has been
  found to have a small impact on the final results by
  Ref.~\cite{McDonald:05}.

Within the range allowed by the data, the effect of $\Lambda$CDM
cosmological parameters and astrophysical parameters on the flux power
spectrum is nearly linear, as can be checked {\it a posteriori} from
the fact that all marginalized probabilities are almost Gaussian.
This justifies the use of a first-order Taylor expansion. The authors
of Ref.~\cite{Viel:05c} checked this explicitly for some parameters,
by performing a second order Taylor expansion or by trying to estimate
the flux power in regions of the parameter space that are very far
from the best-fit with a full hydrodynamical simulation. They usually
found good agreement between the various methods at the percent level
in the simulated flux power.

In the present work, we want to obtain constraints on mixed
$\Lambda$CWDM models. The parameter probability distribution is
strongly non-Gaussian with respect to the extra parameters
characterizing this set-up: the density fraction of warm dark matter,
and its mass (or velocity dispersion). Hence, a simple first-order
Taylor expansion would lead to inaccurate results. In section
\ref{sec:sdss-results}, we extend the method of \cite{Viel:05c} by
interpolating between a grid of fiducial models in this
two-dimensional sub-space, while keeping a simple Taylor expansion in
other parameters. To this end, we performed a grid of 8 additional
simulations (for two values of the masses and four values of the warm
dark matter fraction). In the next section, we will discuss the
robustness of these simulations, especially as far as the issue of
initial velocities is concerned.

\section{Thermal velocities in ICs}
\label{sec:veloc-init-cond}

In N-body or hydrodynamical simulations, the difference between
Initial Conditions (ICs) for CDM and WDM does not reside entirely in a
modification of the power spectrum, but also in non-negligible thermal
velocities for the warm component. For Thermal Relics (TR) assumed to
decouple from thermal equilibrium when they are still relativistic,
and characterized by a Fermi-Dirac distribution with temperature
$T_{\textsc{tr}}$, the average velocity is given by
\begin{equation}
\langle v_{\textsc{tr}} \rangle = \frac{3.151 ~ T_{\textsc{tr}}(z)}{m_{\textsc{tr}}}
= \left( \frac{1+z}{100} \right) \left( \frac{1~\kev}{m_\textsc{tr}} \right)
\left( \frac{T_{\textsc{tr}}}{1~\mathrm{K}} \right) 
\, 8.07 ~\mathrm{km.s^{-1}}.
\label{eq:vtr}
\end{equation}
However, in this paper, we assume that WDM consists in sterile
neutrinos produced through a Non-Resonant Production (NRP) mechanism,
namely, non-resonant oscillations with active
neutrinos~\cite{Dodelson:93,Dolgov:00,Asaka:06c}. In this case the
distribution can be roughly approximated by the renormalized
Fermi-Dirac distribution of Eq.~(\ref{eq:fp}),
corresponding to an average velocity
\begin{equation}
  \label{eq:vnrp}
  \langle v_\dw\rangle = \frac{3.151 T_\nu(z)}{m_\dw} = \frac{3.151
    (4/11)^{1/3} (1+z) T_\textsc{cmb}}{m_\dw} = \left( \frac{1+z}{100}
  \right) \left( \frac{1~\kev}{m_\dw} \right) \, 15.7
  ~\mathrm{km.s^{-1}}.
\end{equation}
If the mass and the relic density $\Omega_\wdm h^2$ are specified, the
parameter $T_\textsc{tr}$ (for TR) or $\chi$ (for NRP) can be inferred
from $(T_\textsc{tr}/T_\nu)^3 = \chi = \Omega_\wdm h^2
(94~\mathrm{eV}/m)$.  Thermal relics and sterile neutrinos sharing the
same mass and density correspond to different initial velocities.  For
instance, assuming that $\Omega_\wdm h^2=0.12$, we find the following
velocity dispersion at $z=99$ in each of the two cases:
\begin{align}
  \label{eq:6}
  m &= 10\kev & \langle v_\dw\rangle &\approx 1.6\;\mathrm{km/sec} & \langle
  v_\textsc{tr}\rangle &\approx 0.16\:\mathrm{km/sec}\notag\\
  m &= 5\kev & \langle v_\dw\rangle &\approx 3.1\;\mathrm{km/sec} & \langle
  v_\textsc{tr}\rangle &\approx 0.41\:\mathrm{km/sec}\\
  m &= 0.5\kev & \langle v_\dw\rangle &\approx 31\;\mathrm{km/sec} & \langle
  v_\textsc{tr}\rangle &\approx 8.8\:\mathrm{km/sec}\notag
\end{align}
These velocities should be compared with the peculiar velocities which
particles acquire when clustering. In most codes, the latter are
initialized using the so-called Zel'dovich
prescription~\cite{Bertschinger:95,Klypin:97,Klypin:00}: so, later, we
will make the distinction between {\it thermal velocities} and {\it
  Zel'dovich velocities}.  Whenever the thermal velocities are
negligible with respect to the Zel'dovich ones, the final result is
expected to be insensitive to the proper implementation of thermal
velocities, which can then be omitted from ICs. In a typical
simulation, the  Zel'dovich velocities are of below or around  20~km/s
at $z = 99$. Hence, for the NRP case, we
roughly expect. Hence, for the NRP case, we roughly expect that
thermal velocities are unimportant for $m \geq 5$~keV.

This can be checked explicitly.  Our numerical simulations are performed with
the \textsc{gadget-ii} code~\cite{Springel:05a}, extended in order to compute
the Ly-$\alpha$ flux power spectrum~\cite{Viel:04}. First, we address the case
of pure $\Lambda$WDM models with $\Omega_\cdm=0$. We fix initial conditions at
$z=99$.  The initial power spectrum is computed with \textsc{camb}, and for
each warm particle a random thermal velocity is eventually added to the
Zel'dovich velocity.  Thermal velocities are distributed with a Fermi-Dirac
probability.
% Notice, however, that a particle in the simulated ICs represents a
% region with size $l = L_{box}/N^{1/3}$, where $N$ is the number of
% particles. 
% The assignment of a Fermi-Dirac thermal velocity to the
% particle is justified provided that its size is smaller than the smallest
% scale probed by our data set, 
% and larger than the extra distance traveled by the blob during the
% whole simulation due to thermal velocities.  In our simulation, we
% take $N=400^3=6.4\times 10^7$ particles in a box of comoving size
% $L_{box} = 60\mpc$/h, so each resolution element in the ICs measures
% $l_{res}\sim 0.15\mpc$ in comoving units, while the smallest scale
% probed by our Ly-$\alpha$ flux power spectrum corresponds to a
% comoving wavelength of the order of $1 \mpc$. In addition, the extra
% distance travelled by the blob can be estimated conservatively by
% multiplying the velocity $v \sim 10$~km/s by the amount of time
% between $z=99$ and $z=2$, $t \sim 3$~Gyr: this gives 0.04~Mpc, which
% remains smaller than the blob size. Hence, our approach is justified.

% A similar way to include velocities in the WDM simulations
% was used e.g. in~\cite{Colin:07}.  Note that the simulated flux power
% has been corrected for resolution using exactly the same procedure
% described in \cite{Viel:07}, so even if most of the simulations have
% been ran at somewhat low resolution we corrected for it using the
% higher resolution runs.

Notice, however, that a particle in the simulated ICs represents a
region with size $l = L_{box}/N^{1/3}$ (where $N$ is the number of
particles) .  Therefore, a huge amount of DM particles are represented
as one body in Â N-body simulations. Formally, the velocity of such a
collection of particles is zero, as the velocities of real DM
particles are random and they average to zero.  The assignment of a
Fermi-Dirac thermal velocity to such a simulation particle is
justified provided that $l$ is smaller than the smallest scale probed
by our data set, and larger than the extra distance traveled by the
particle, due to thermal velocities, when it moves along Zel'dovich
trajectories.  In our simulation, we take $N=400^3=6.4\times 10^7$
particles in a box of comoving size $L_{box} = 60\mpc$/h, so each
resolution element in the ICs measures $l\sim 0.15\mpc$ in comoving
units, while the smallest scale probed by our Ly-$\alpha$ flux power
spectrum corresponds to a comoving wavelength of the order of $1
\mpc$.  In addition, the extra distance travelled by the particle can
be estimated conservatively by multiplying the velocity $v \sim
10$~km/s by the amount of time between $z=99$ and $z=2$, $t \sim
3$~Gyr: this gives 0.04~Mpc,which is smaller than the resolution
element probed by the particle in the ICs. Hence, our approach is
justified.  A similar way to include velocities in the WDM simulations
was used e.g. in~\cite{Colin:07}.  Note that the simulated flux power
has been corrected for resolution using exactly the same procedure
described in \cite{Viel:07}, so even if most of the simulations have
been ran at somewhat low resolution we corrected for it using the
higher resolution runs.

\begin{figure}[t]
  \centering
  \includegraphics[width=\textwidth]{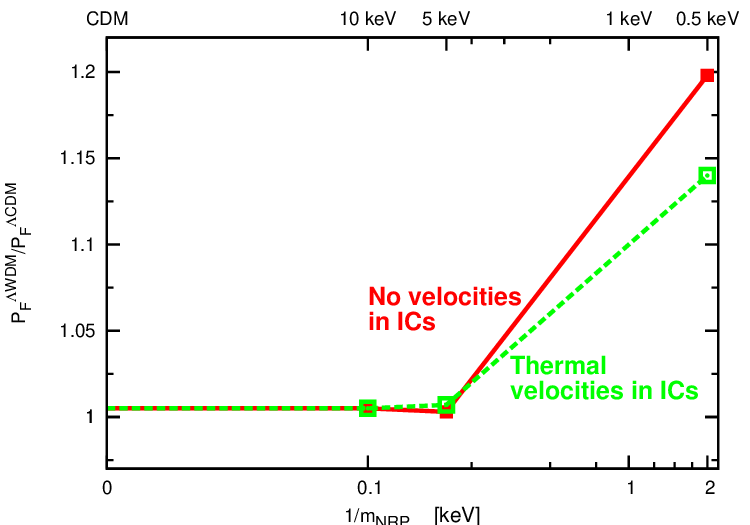} %
  \caption{Dependence of the flux PS of $\Lambda$WDM models (divided
    by that of $\Lambda$CDM) on thermal velocities in hydrodynamical
    simulations. The points show $P_F^{\Lambda
      \mathrm{WDM}}/P_F^{\Lambda \mathrm{CDM}}$ at $k= 0.14 \:
    h/\mathrm{Mpc}$ and $z=2.2$, extracted from \textsc{GADGET-II}
    simulations for three different masses, with and without thermal
    velocities in the initial conditions. The lines are the results of
    linear interpolation between these points.}%
  \label{fig:wdm-velocity}
\end{figure}

In Fig.~\ref{fig:wdm-velocity}, we show the ratio of the $\Lambda$WDM
flux power spectrum to the $\Lambda$CDM one at $k= 0.14 \: h/\mathrm{Mpc}$
and $z=2.2$
for 3 masses ($0.5, 5, 10\kev$), with and without thermal velocities
in ICs.  One can see that for masses above $\sim
5\kev$, the influence of velocities is below the $1\%$ level, which
is smaller than the error bars of the SDSS data points (in case of the
VHS dataset, the influence of velocities is unimportant also for
smaller masses due to the larger error bars).  For smaller masses,
including thermal velocities in the initial conditions becomes
essential.

This suggest that also in the case of a mixed $\Lambda$CWDM, we can
perform simulations without thermal velocities provided that the mass
of the WDM NRP component is above $\sim 5\kev$. This simplifies
considerably the issue of initial conditions, because in this case we
can treat the mixed CDM+WDM components as a single cold fluid with
negligible thermal velocities, and a proper initial power spectrum
(equal to the total linear matter power spectrum at high redshift,
as computed by {\sc camb}).

\section{Results for VHS data}
\label{sec:results-vhs}

\begin{figure}[tp]
  \centering
  \includegraphics[width=\textwidth]{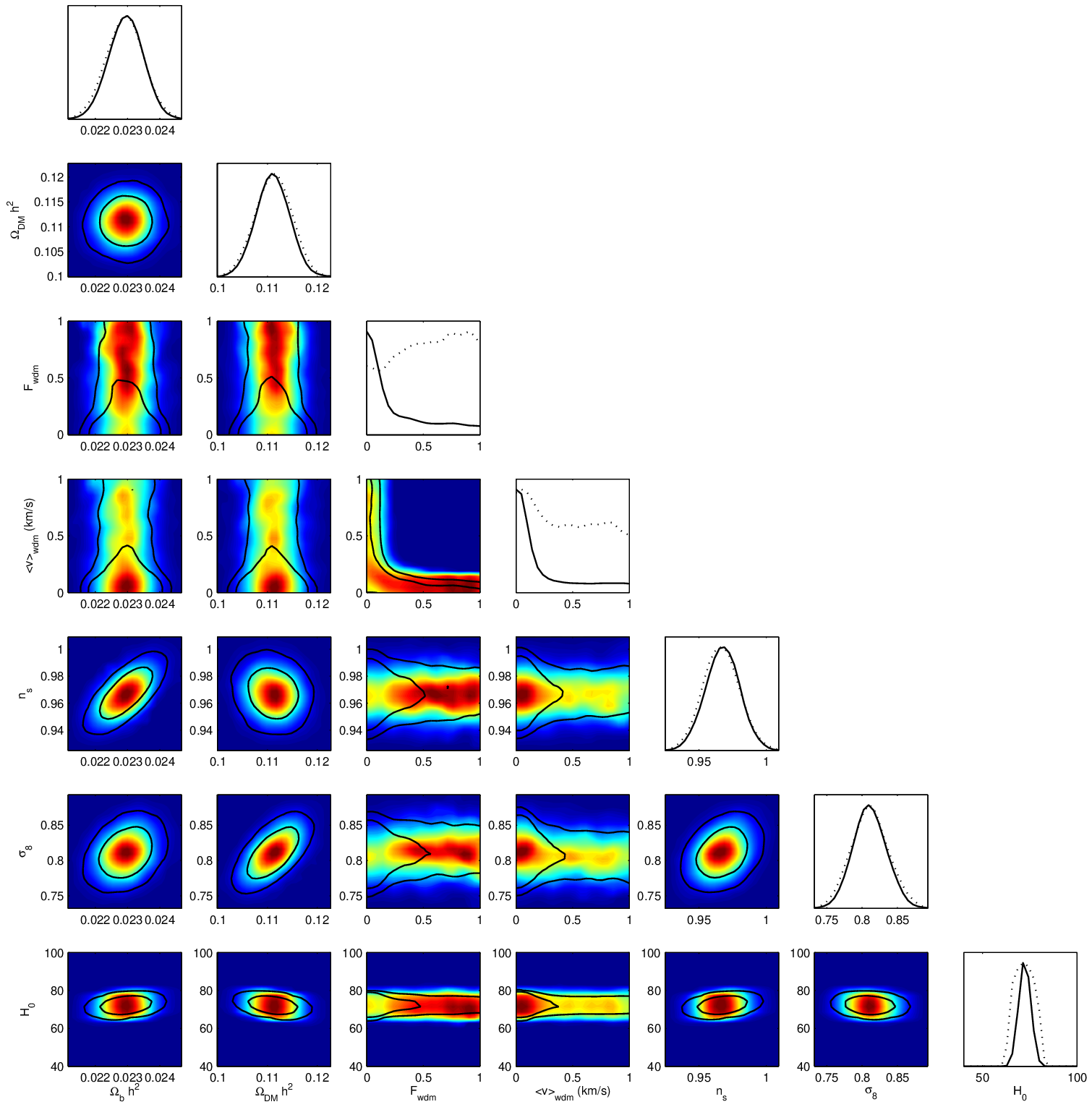}
  \caption{Preferred region for pairs of cosmological parameters in
    the $\Lambda$CWDM model, using VHS, WMAP5 and other cosmological
    data sets. The black lines represent the $68\%$CL and $95\%$CL
    contours of the two-dimensional probability marginalized over
    other parameters (as defined in the Bayesian approach), while the
    colors show the likelihood averaged over other parameters.  The
    first five parameters $\Omega_\bary h^2$, $\Omega_\dm h^2$,
    $F_\wdm$, $\langle v\rangle_\wdm$, $n_s$ have flat priors over the
    displayed range.  Our analysis also assumes flat priors on the
    logarithm of the primordial spectrum amplitude and on the angular
    diameter of the sound horizon, but instead of these parameters we
    show here $\sigma_8$ and $H_0$, which have a more straightforward
    interpretation. In this plot, we do not show the reionization
    optical depth $\tau$ and all the nuisance parameters entering into
    the analysis. The one-dimensional marginalized probabilities are
    displayed along the diagonal.}
  \label{fig:vhs}
\end{figure}

\begin{figure}[t]
  \centering
  \includegraphics[width=\textwidth]{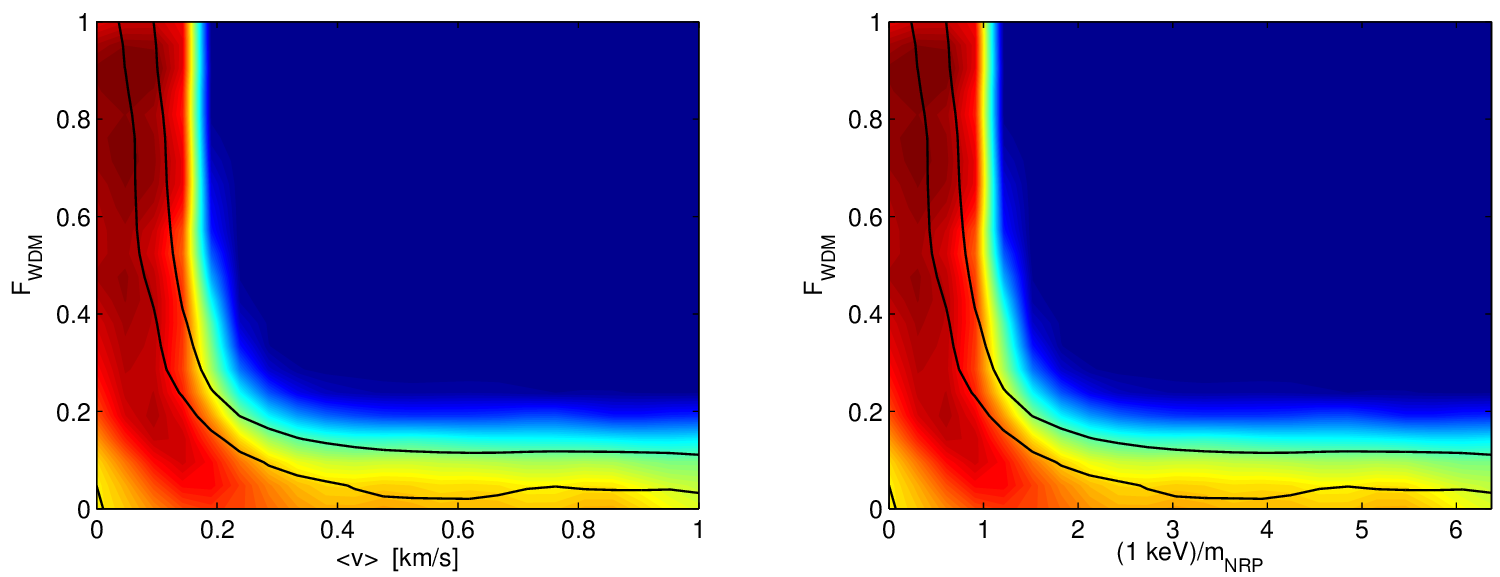}
  \caption{{\it (Left)}~Preferred region for the parameters ($\Fwdm,
    \langle v\rangle_\wdm)$ of the $\Lambda$CWDM model, using VHS,
    WMAP5 and other cosmological data sets. The black lines represent
    the $68\%$CL and $95\%$CL contours of the two-dimensional
    probability marginalized over other parameters (as defined in the
    Bayesian approach), while the colors show the likelihood averaged
    over other parameters.  {\it (Right)}~Same plot with the
    horizontal axis translated in terms of $(1 \mathrm{keV} / m_\dw) =
    \langle v\rangle_\wdm / (157~\mathrm{m/s})$, see equation
    (\ref{eq:vnrp}).  }
  \label{fig:vhs_2D}
\end{figure}

We performed a Bayesian parameter estimation for the $\Lambda$CWDM
model, using the following data set: VHS \lya data;
WMAP5~\cite{Dunkley:2008ie} and small-scale CMB experiments
(ACBAR~\cite{Kuo:2002ua}, CBI~\cite{Readhead:2004gy},
Boomerang~\cite{Piacentini:2005yq,Jones:2005yb,Montroy:2005yx,MacTavish:2005yk});
galaxy power spectrum from the SDSS LRGs~\cite{Tegmark:2006az};
supernovae from the SNLS~\cite{Astier:2005qq}.  We take flat priors on
six cosmological parameters describing the $\Lambda$CDM model
($\Omega_\dm h^2,\Omega_\bary h^2,\theta,\tau,n_s,A_s$)\footnote{See
e.g.~\cite{Lewis:02}
and~\url{http://cosmologist.info/cosmomc/readme.html} for explanation
of this parameter choice.}, on two extra parameters describing the
warm sector (see below), and we marginalize over nuisance parameters
associated with the data sets, in the way implemented in the public
version of {\sc CosmoMC} (compiled with the file {\tt lya.f90}).  The
two extra parameters are chosen to be:
\begin{itemize}
\item first, the warm fraction of dark matter,
\begin{equation}
\Fwdm \equiv \frac{\Omega_\wdm}{\Omega_\cdm + \Omega_\wdm},
\end{equation}
which differs from the quantity $\fwdm$ defined in Eq.~(\ref{eq:3}) by
$(\Omega_\dm / \Omega_M)$ since the denominator accounts for dark matter
instead of total matter; indeed, in the analytical solutions of
section~\ref{sec:plateau-cwdm-models}, $\fwdm$ is the quantity which appears
naturally; in a parameter extraction, $\Fwdm$ is more convenient because of
its clear interpretation: it varies from $\Fwdm=0$ for pure cold to $\Fwdm=1$
for pure warm dark matter.  Instead, $\fwdm=1$ would describe a model with
negligible CDM {\it and} baryonic components, providing very bad fits to the
data, and an eventual upper limit on $\fwdm$ would not be straightforward to
interpret physically.
\item the average velocity of warm particles today in km/s: $\langle
  v\rangle_\wdm$.  This parameter is convenient because, unlike the particle
  mass, it has a direct physical effect on the power spectrum; hence our
  results apply to both NRP and TR models (and presumably to a much wider
  class of WDM models).  Bounds on $\langle v\rangle_\wdm$ can be translated
  immediately in terms of $m_{\dw}$ using Eqs.~(\ref{eq:vnrp}); limits on
  $m_\textsc{tr}$ are not so straightforward to obtain since they depends on
  the posterior distribution of $\langle v\rangle_\wdm$, $\Fwdm$ and
  $\Omega_\dm$.
\end{itemize}
Note that varying $\Fwdm$ and $\langle v\rangle_\wdm$ is equivalent to tuning
respectively the size and the location of the step-like suppression in the
matter power spectrum (at least at the linear level), as shown in
Section~\ref{sec:plateau-cwdm-models}.

The main goal of the various MCMC runs performed in this work is to
derive some (Bayesian) credible intervals\footnote{See the Appendix A
  for a comparison of Bayesian {\it credible intervals} and
  frequentist {\it confidence limits}.} for the parameter $\langle
v\rangle_\wdm$, starting from a flat prior on this parameter in the
range $0<\langle v\rangle_\wdm < \infty$. These results can be
immediately translated in credible intervals for $m_\dw^{-1}$, using
the relation $(1~\mathrm{keV})/m_\dw = \langle v\rangle_\wdm /
(0.157~\mathrm{km/s})$, see equation (\ref{eq:vnrp}) with $z=0$.  It
would not be convenient to derive directly some credible intervals on
$m_\dw$ itself, since the $\Lambda$CDM limit (which is perfectly
allowed by the data) corresponds to $m_\dw \longrightarrow
\infty$. However, each 95\% Confidence Level (CL) upper bound on
$\langle v\rangle_\wdm$ can be expressed as a lower bound on the mass,
that we will denote as $m_\dw^{95}$. We warn the reader that this
bound should be interpreted with care, since we are not using a flat
prior on the mass, but rather on the velocity, i.e. on the inverse
mass. Hence the quantity $m_\dw^{95}$ represents a Bayesian 95\%~CL
upper limit on the velocity expressed in a useful way.

We first ran {\sc CosmoMC} with $\Fwdm$ fixed to one, in order to
update bounds on the pure $\Lambda$WDM model. We found a credible
interval $0 < \langle v\rangle_\wdm < 0.10 \:\mathrm{km/s}$ (95\%~CL),
which can be translated into the lower limit $m_\dw^{95}= 1.6$~keV.
In reference~\cite{Viel:05} (which also used VHS \lya data, but WMAP1
instead of WMAP5), a stronger bound
$m_\dw^{95} = 2.0
\kev$ was found. 
This is surprising at first sight, since
the CMB temperature and polarization spectra are better measured by
WMAP5. 
In addition, WMAP5 favors the lowest range of $n_s$ and
  $\sigma_8$ values allowed by WMAP1, so one could expect that a
  cut-off in the power spectrum due to WDM is even more unlikely,
  leading to stronger bounds.  Actually, the constraints on this
  cut-off come entirely from the VHS data: so, for $\Lambda$WDM
  models, the real expectation is that the mass bounds should remain
  almost insensitive to variations in the CMB data.
Still, a possible explanation of this $20\%$ discrepancy is that for
$\Lambda$WDM (and $\Lambda$CWDM) models, the convergence of MCMCs
requires a huge amount of samples. In this work, we deliberately
accumulated a very large number of chain points, and checked
thoroughly that our chains are fully converged\footnote{
  This particular $\Lambda$WDM run is based on 6 independent chains.
  After accumulating a total of $\sim 30\,000$ chain points, we
  reached a convergence criterion $(R-1)\sim 0.02$ (based on the
  second half of the full chains, see Ref.~\cite{Lewis:02} for a
  definition of $R$), and a bound $m_\dw^{95} = 2.0 \kev$. However, it
  is only after accumulating $\sim100\,000$ chain points that the
  bound stabilized to $m_\dw^{95} = 1.6 \kev$, with $(R-1)\sim
  0.007$. We know that this bound is accurate because we monitored the
  evolution of the mass bound versus execution time, and checked that
  its variations remain negligible (less than 5\%) during the second
  half of the run; also, we checked that if we vary the fraction of
  points to remove at the beginning of each chain in the range
  20-80\%, or if we remove one of the chains, the bound remains
  precisely equal to $1.6 \kev$. In the next section on SDSS results,
  we will use {\sc CosmoMC} with a more convenient convergence
  criterion than $R$ in order to establish the precision and stability
  of all 95\%~CL credible limits}. The older result might have been
affected by poorer chain convergence.

Notice that our $95\%$ CL credible limit on $m_\dw$ in $\Lambda$WDM
models is very close to that obtained from phase-space density
arguments~\cite{Tremaine:79} applied to newly discovered dwarf
spheroidal satellites of the Milky Way~\cite{Boyarsky:08a}, which give
$m_\dw > 1.77\kev$ (see also~\cite{Gorbunov:08b}).

We then ran {\sc CosmoMC} for the full $\Lambda$CWDM model, with one
more parameter $\Fwdm$. Our results are summarized in
Figure~\ref{fig:vhs}. The plots on the diagonal show the probability
of each parameter.  The plots off-diagonal show the (Bayesian)
credible limits (68\%~CL or 95\%~CL) on pairs of parameters.

Figure~\ref{fig:vhs_2D} focuses on the two-dimensional probability
distribution in the plane $(\Fwdm, \langle v\rangle_\wdm)$ or $(\Fwdm,
(1 \mathrm{keV}) / m_\dw)$. As $\Fwdm$ approaches one (pure WDM
limit), the 95\%~CL contour reaches $\langle v\rangle_{\wdm} \sim 0.10
\:\mathrm{km/s}$, which corresponds to a lower bound on the mass
$m_\dw^{95} = 1.6 \kev$ (coincidentally, this is identical to the
bound derived from the one-dimensional probability of $\langle
v\rangle_{\wdm}$ in the $\Lambda$WDM case).  In the opposite limit, as
$\langle v \rangle_\wdm$ grows, we qualitatively expect that the
contour will tend towards a constant value, since below a particular
value of $\Fwdm$ the impact of the WDM component on the linear power
spectrum will become too small for being detectable by the VHS data,
given the error bars and calibration uncertainty of this data.  We
find indeed that the 95\%CL contour reaches asymptotically $\Fwdm
\simeq 0.1$: that is, an admixture of about 10\% of WDM with an
arbitrary value of the mass within our prior bound ($m_\dw \geq
160\ev$) is compatible with the VHS \lya data\footnote{Notice,
  however, that a $\Lambda$CWDM model with $10\%$ of WDM and $m_\dw
  \simeq 160\ev$ is ruled out by phase-space density
  considerations~\cite{Boyarsky:08a}.  Indeed the bound $m_\dw >
  1.77\kev$ presented in~\cite{Boyarsky:08a} scales as
  $f_\wdm^{1/3}$. Therefore when reducing the fraction of WDM from
  100\% to 10\%, it diminishes by 50\% only and becomes $m_\dw >
  0.89\kev$.}.

\section{SDSS results}
\label{sec:sdss-results}

We now consider the bounds that can be obtained by combining two data
sets: WMAP5~\cite{Dunkley:2008ie} and the SDSS \lya data introduced in
section \ref{sec:data:sdss}. We did not include any other
experimental results in this section.

\subsection{Pure WDM model}

Before applying the method described in section \ref{sec:data:sdss} to
the $\Lambda$CWDM model, we updated the analysis of \cite{Viel:06}
for a pure $\Lambda$WDM model. 
The new ingredients in this work are
\begin{inparaenum}[\em(i)]
\item the use of WMAP5 rather than WMAP3 data;
\item the suggestion of Ref.~\cite{Bolton:07a} to marginalize the
  astrophysical parameter $\gamma$ over a more conservative range
  (recall $\gamma$ describes the relation between the local temperature of
  the intergalactic medium and its local overdensity, $T=T_{0}
  (1+\delta)^{\gamma-1}$).
\end{inparaenum}
Apart from these new inputs, we performed several tests concerning the
validity and robustness of the method, trying different schemes of
interpolation between the points where the hydrodynamical simulations were
performed, implementing or not thermal velocities in the initial conditions,
testing extensively the convergence of MCMCs, and comparing Bayesian and
frequentist limits.  For these tests we performed \textsc{GADGET-II}
simulations for three different values of the mass, $m_{\dw}=0.5, 5, 10\kev$
(instead of a single one, used in~\cite{Viel:06}).

\subsubsection{Bayesian credible interval}
\label{sec:depend-wdm-results}

\begin{figure}[t]
  \centering
  \includegraphics[width=\textwidth]{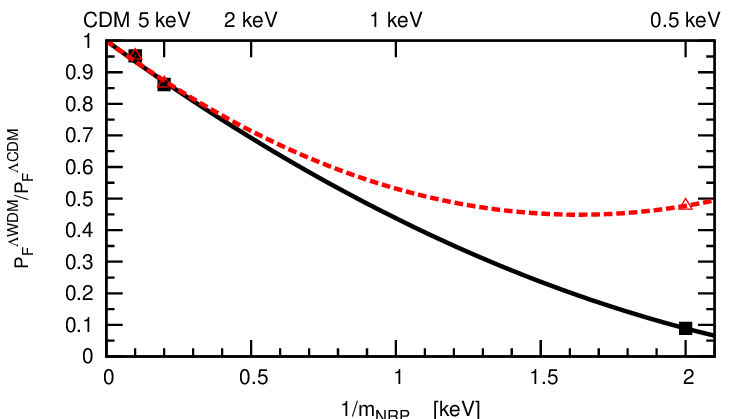}
  \caption{Dependence of the interpolated flux PS of $\Lambda$WDM
    models (divided by that of $\Lambda$CDM) on thermal velocities in
    ICs, here for $k = 0.018 s/km$ and $z = 4.2$. The points are
    obtained with hydrodynamical simulations; the curves correspond to
    a global quadratic fit of these points.}
  \label{fig:wdm-fit}
\end{figure}

\begin{figure}[t]
  \centering
  \includegraphics[width=0.45\textwidth]{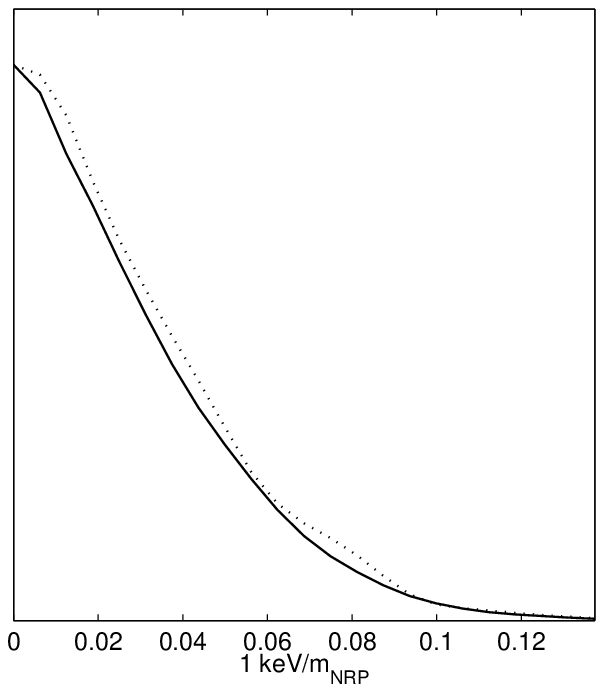}
  \caption{Probability distribution for the parameter $1 \,
    \mathrm{keV}/m_\dw$ (which is identical to $\langle v \rangle_\wdm /
    (157~\mathrm{m/s})$) in the pure $\Lambda$WDM case, using WMAP5 and SDSS
    \lya data, and starting from a flat prior on this parameter; the solid
    line shows the marginalized posterior, while the dashed line shows for
    comparison the average likelihood. The normalization of the y-axis (the
    probability) is arbitrary.}
  \label{fig:m-prob}
\end{figure}

It is important to check whether including or not thermal velocities
into the ICs has an impact on the WDM mass bounds. The analysis of
section \ref{sec:veloc-init-cond} shows that thermal velocities are
expected to change significantly the simulated flux PS at least for
$m_{\dw} \lesssim 0.5\kev$, while the effect of velocities remains
very small at least for $m_{\dw} \gtrsim 5\kev$. Since previous
works~\cite{Seljak:06,Viel:06} obtained a (Bayesian) lower bound of
the order of $m_\dw^{95} = 10\kev$ (95\% CL), one expects the influence
of thermal velocities in ICs onto the lower mass bound to be small. To
analyze this issue in details we performed a number of tests.

  In our analysis, the dependence of $P_F^{\Lambda\wdm}(k,z)$ on
  $m_\dw$ for each value of $(k,z)$ is obtained by interpolation,
  following a global quadratic fit to the four points for which we
  have exact results ($m_\dw = 10, 5, 0.5$~keV, and the pure
  $\Lambda$CDM case $m_\dw=\infty$). The result of such a fit is shown
  in figure \ref{fig:wdm-fit} for particular values of $k$ and $z$.
  For $m_\dw < 5$~keV, the flux PS is strongly affected by the
  presence or absence of thermal velocities in the hydrodynamical
  simulations with $m_\dw = 0.5$~keV, even with some level of
  arbitrariness since various interpolation schemes could be used
  between $m_\dw = 0.5$~keV and $m_\dw = 5$~keV. Even the values of
  $P_F$ for $m_\dw\ge 5 \, \kev$ are slightly affected at the per cent
  level by the difference between the two interpolations\footnote{ We
    have also compared different ways of interpolating the flux PS
    between the four points at our disposal for each $(k,z)$ and found
    that above $m_\wdm \gtrsim 5\kev$, the difference between the
    various interpolation schemes is at most of the order of several
    per cents for each $P_F^{\Lambda \cwdm}(k,z)$, which is comparable
    with the error bars on the data points of the SDSS set. The
    resulting errors in $m_\dw^{95}$ are less than 10\%.}.  Thus, the
  dependence on velocities in ICs can potentially affect the pure WDM
  result, which depends on the flux spectrum variation for $m_\dw
  \gtrsim 10\kev$.

We have performed MCMC runs for both types of simulated flux PS (with and
without thermal velocities in ICs), using a Bayesian top-hat priors on the
parameter $0 \leq \langle v \rangle_\wdm \leq 0.3$~km/s (corresponding to
$m_\dw \ge 0.5\kev$). At the 95\% CL, we found $0 \leq \langle v \rangle_\wdm
\leq 11.6$~m/s (with velocities) or $0 \leq \langle v \rangle_\wdm \leq
11.1$~m/s (without velocities), corresponding to the bounds $m_\dw^{95} =
13.5$~keV (with velocities) or $m_\dw^{95} = 14.1$~keV (without velocities). In
figure \ref{fig:m-prob}, we show the marginalized probability distribution for
the variable $1~\mathrm{keV}/m_\dw$, which is identical to $\langle v
\rangle_\wdm / (157~\mathrm{m/s})$ (see equation (\ref{eq:vnrp}) at $z=0$).

As any credible interval limits obtained with an MCMC method, the
above values of $m_\dw^{95}$ have some uncertainty related to the fact
that there is no clear and unique criterion of convergence for
MCMCs~\cite{Lewis:02}. It is always difficult to decide when the
bounds are considered to be stable and when to stop \textsc{CosmoMC}.
To quantify this ambiguity, we used some convergence criteria
implemented into {\sc CosmoMC}: {\tt \small MPI\_Limit\_Converge\_Err
  = 0.3} and {\tt \small MPI\_Limit\_Converge = 0.025}, which mean
that the run stopped when the 95\%~CL credible limits were estimated
to be accurate at the level of 30\% of the standard deviation for each
parameter.  In the case of $\langle v \rangle_\wdm$, the standard
deviation was found to be $\sigma=3.6$~m/s, so the error on the bounds
coming form the finite degree of convergence of the chains was of the
order of $\sim 1$~m/s. This means that our bounds for the velocity and
for the mass are only accurate up to $\pm$~10\%.  Indeed, this
reflects the typical variations that one observes when varying the
fraction of the chains included in the analysis, when removing some of
the chains, etc.  Including this source of error, the two bounds
obtained with and without thermal velocities are found to be
compatible with each other, and 
are summarized by $m_\dw^{95} = 12.1$~keV. This final result
for the $\Lambda$WDM case, which takes into account the systematic
uncertainty of the MCMC method, depends neither on the interpolation
scheme, nor on the inclusion thermal velocities in ICs.

This discussion suggests that in the mixed $\Lambda$CWDM case, we should use a
top-hat prior $0 \leq \langle v \rangle_\wdm \leq 0.03$~km/s (corresponding to
$m_\wdm \ge 5\kev$) in order to obtain conservative bounds; in this way, our
final results will not depend on the treatment of thermal velocities; and more
importantly, they will not be affected by the interpolation scheme, which
would remain somewhat arbitrary even if the thermal velocities were correctly
implemented in {\sc gadget}. In order to obtain robust results below $m_\wdm
\sim 5\kev$, one should perform more hydrodynamical simulations for
intermediate values of the mass, and perform extensive tests of the
convergence of {\sc gadget} results w.r.t thermal velocity effects. This
analysis is beyond the scope of the present work.

\subsubsection{Comparison with frequentist approach}
\label{sec:runs-fixed-param}

\begin{table}[t]
\begin{tabular}{l|c|c|c|l}
  Run & Â $-\log(\mathcal{L}_\text{best})$ &
  $2\Delta\log(\mathcal{L}_\text{best})$ & 
  \# params. & Comment\\
  \hline
%  WDM, $m_\dw$ varying Â & 1404.2 & 0.0 & 34 & $m_{95} = 13.5\kev$ \\
%  \hline
  CDM Â ($\Leftrightarrow m_\dw = \infty$) Â  Â  Â  & 1404.0 & 0.0 &
  33 &
  \\
  WDM, $m_\dw = 13.5\kev$ Â & 1404.6 & 1.2 & 33 &$m_\dw = m_\dw^{95}$\\ %&
  WDM, $m_\dw = 11.2\kev$ Â & 1405.2 & 2.4 & 33 &$m_\dw = m_\dw^{95}/1.2$\\
  WDM, $m_\dw = 9.0\kev$ Â & 1406.3 & 4.6 & 33 &$m_\dw = m_\dw^{95}/1.5$\\ 
  WDM, $m_\dw = 6.75\kev$ Â & 1412.0 & 16.0 & 33 &$m_\dw = m_\dw^{95}/2.0$\\
  \hline
\end{tabular}
\caption{Dependence of the best likelihood in our MCMCs (for SDSS
  Lyman-$\alpha$ and WMAP5 data) on the mass $m_\dw$. The five lines
  correspond to distinct sets of chains, obtained by running {\sc
    CosmoMC} with \emph{fixed} values of $m_\dw$: CDM limit
  (equivalent to $m_\dw=\infty$), and $m_\dw$ equal to $m_\dw^{95}$
  divided by 1, 1.2, 1.5 and 2.0, where $m_\dw^{95}$ is the mass value
  corresponding to the Bayesian 95\% CL upper bound on $\langle v
  \rangle_\wdm$. Notice that these five cases have the same number of
  degrees of freedom (data points minus parameters).}
\label{tab:sdss}
\end{table}

\begin{figure}[ht]
  \centering
  \includegraphics[width=\textwidth]{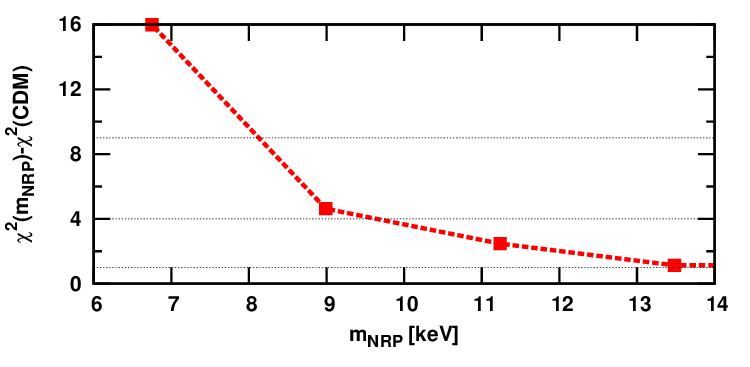}
  \caption{Dependence of $\Delta \chi^2 = - 2 \log \mathcal{L}$ on
    the WDM mass $m_\dw$. The values are taken from Table~\ref{tab:chi2}, and
    the dashed line linearly interpolates between the points.}
  \label{fig:fixed-mass}
\end{figure}

In the previous section, we found that the 95\%~CL credible interval
for $\langle v \rangle_\wdm$ corresponds to a lower limit on the mass
$m_\dw^{95} = 13.5$~keV (not including the additional uncertainty
related to MCMC convergence issues, which led us to lower the
  final bound to $m_\dw^{95}=12.1$~keV).  It is well known that the
Bayesian approach finds the \emph{most probable} parameter values
given the data (see Appendix~\ref{sec:confidence-credible} for a short
introduction to Bayesian and frequentist analyses). Therefore the
Bayesian $95\%$~CL limits on each parameter border the region
containing 95\% of the parameter probability (marginalized over other
parameters).  We are also interested in the answer to a different
question: whether for a particular value of $m_\dw$ there exists
\emph{at least one model} which fits the data well enough compared to
pure CDM (regardless of \emph{how many models} fit the data equally
well). Addressing this question is particularly useful for {\it
  excluding} conservatively some values of the mass, since it also
tells beyond which value {\it there is not a single model} providing a
good fit to the data.

If the likelihood of the $\Lambda$WDM model
was a perfect multivariate Gaussian, we
would expect that $m_\dw^{95}$ is the value of the mass for which the
maximum likelihood ($\max[\CL|_{m=m_\dw^{95}}]$) is equal to the absolute
maximum ($\max[\CL]$) times $e^{-2}$. In the general case, this is not
necessarily the case, and the frequentist bounds can differ
significantly from the Bayesian ones.

To estimate the frequentist confidence limits on $m_\dw$, we performed a
number of {\sc CosmoMC} runs with fixed values of the mass. In
Table~\ref{tab:sdss}, we quote the maximum likelihood in each set of chains.
Variations in the logarithm $- 2 \log(\CL_\text{best})$ (which would be
$\chi^2$-distributed in the limit of a multivariate Gaussian) can be
interpreted as variations in the goodness-of-fit for each value of the mass.
By looking at Table~\ref{tab:sdss}, one sees that the best-fit model for
$m_\dw = 13.5\kev$ fits the data almost as well as the pure CDM model (the
differences $\Delta \chi^2=1.2$ is not statistically significant).

As mentioned in Appendix \ref{sec:confidence-credible}, the value of the mass
for which $-2\log(\CL_\text{best})$ worsens by $\Delta\chi^2=4$ (resp.
$\Delta\chi^2=9$) is an approximation for the 95\%CL (resp. 99.7\%CL) lower
bound for $m_\dw$, conventionally quoted as the $2\sigma$ (resp. $3\sigma$
bound) in many data analyses based on the frequentist approach.  In our case,
this happens for $m_\dw \simeq 9.5\kev$ ($2\sigma$ bound) and $m_\dw \simeq
8\kev$ ($3\sigma$ bound) using linear interpolation. Hence, $m_\dw \simeq 8\kev$
appears to be a robust $3\sigma$ lower bound on the WDM mass (NRP case) in the
pure $\Lambda$WDM case.

The above derivation of frequentist bounds could be improved by
studying more values of the mass, finding the maximum likelihood with
higher accuracy, and computing the exact values of $\Delta \chi^2$
corresponding to 95\% and 99.7\% bounds (for non-Gaussian
distributions, these are not exactly equal to 4 and 9). However, we
believe that our treatment is sufficient for concluding that values
$m_\dw \gtrsim 8$~keV are allowed by present data at the
3$\sigma$ level.

\subsection{SDSS results for CWDM}
\label{sec:cwdm-fixed}

In order to analyze the full $\Lambda$CWDM model, we performed extra
GADGET-II simulations for a grid of points with $m_\dw=5,10 \, \kev$
and $\Fwdm=0.2,0.5,0.9$. We have seen that for such values of the
mass, thermal velocities in the ICs can be neglected. Hence, we
treated the cold plus hot mixture as a single species with no thermal
velocities, distributed initially according to the appropriate linear
power spectrum computed by CAMB at $z=99$.  Including $\Lambda$CDM and
$\Lambda$WDM models on the edges, we have a total grid of $3 \times 5$
models with $m_\dw=5 \kev, 10 \kev, \infty$ and $\Fwdm=0, 0.2,0.5,0.9,
1$. For each pair $(k,z)$, we find the value of $P_F^{\lcwdm}$ at any
given point $(m_\dw, \Fwdm)$ by bilinearly interpolating within this grid.

\subsubsection{Bayesian credible region}

\begin{figure}[tp]
  \centering
  \includegraphics[width=\textwidth]{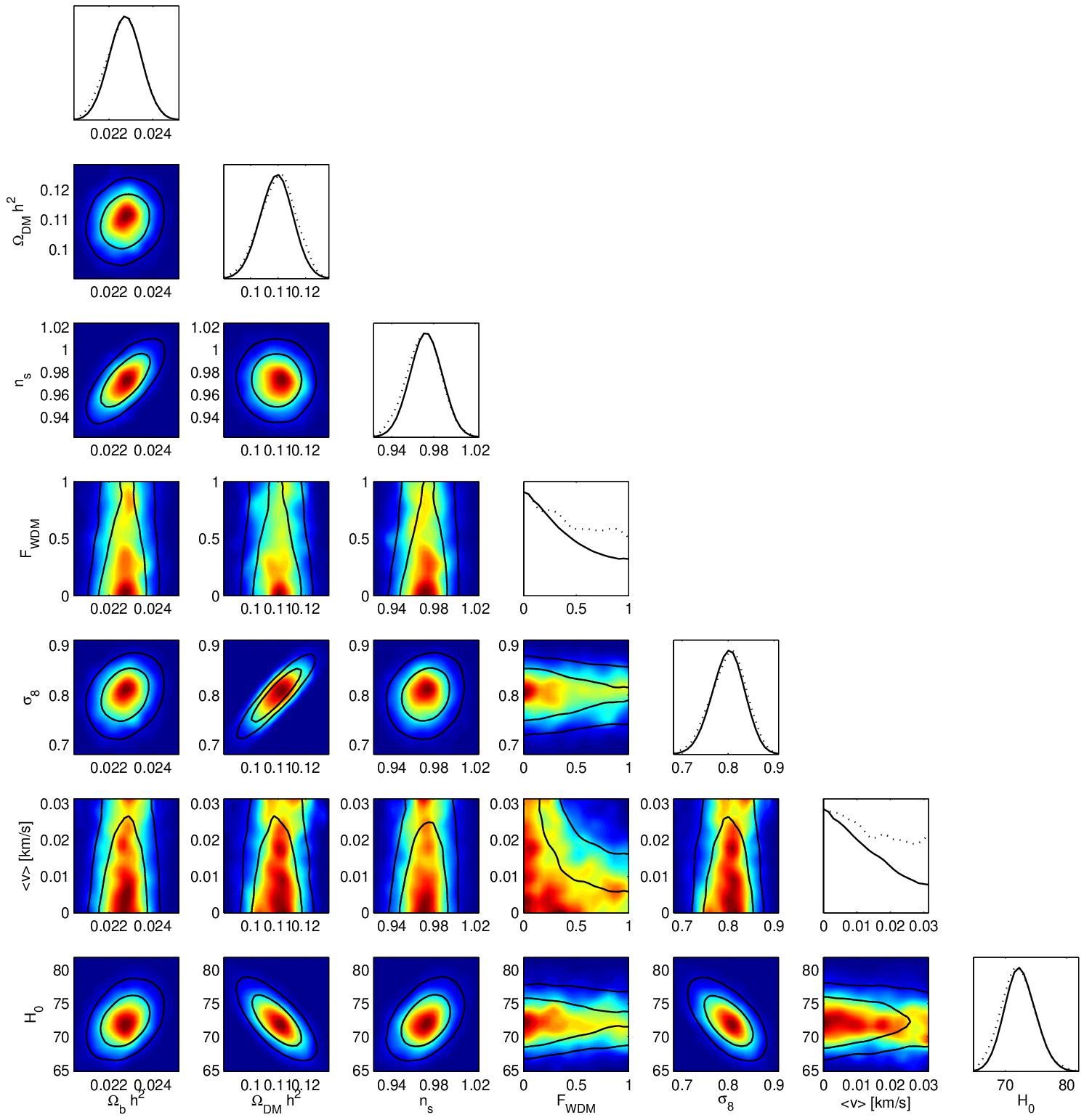}
  \caption{Preferred region for pairs of cosmological parameters in
    the $\Lambda$CWDM model, using SDSS \lya and WMAP5 data sets. The
    black lines represent the $68\%$CL and $95\%$CL contours of the
    two-dimensional probability marginalized over other parameters (as
    defined in the Bayesian approach), while the colors show the
    likelihood averaged over other parameters.  The five parameters
    $\Omega_\bary h^2$, $\Omega_\dm h^2$, $n_s$, $F_\wdm$, $\langle
    v\rangle_\wdm$ have flat priors over the displayed range.  Our
    analysis also assumes flat priors on the logarithm of the
    primordial spectrum amplitude and on the angular diameter of the
    sound horizon, but instead of these parameters we show here
    $\sigma_8$ and $H_0$, which have a more straightforward
    interpretation. In this plot, we do not show the reionization
    optical depth $\tau$ and all the astrophysical or nuisance
    parameters entering into the analysis. The one-dimensional
    marginalized probabilities are displayed along the diagonal.}
  \label{fig:cwdm}
\end{figure}

\begin{figure}[tp]
  \centering
  \includegraphics[width=0.48\textwidth]{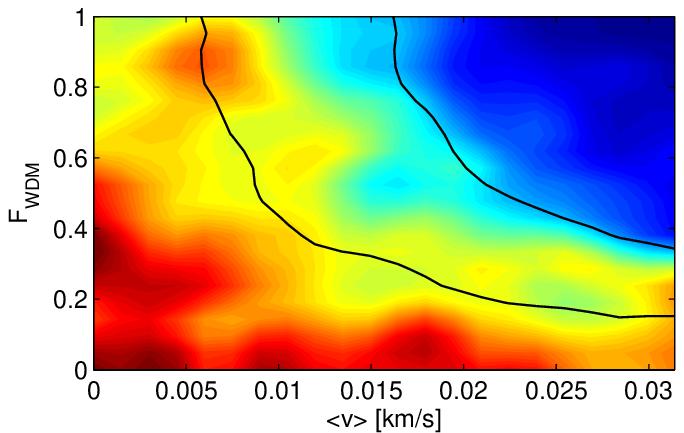}
  \includegraphics[width=0.48\textwidth]{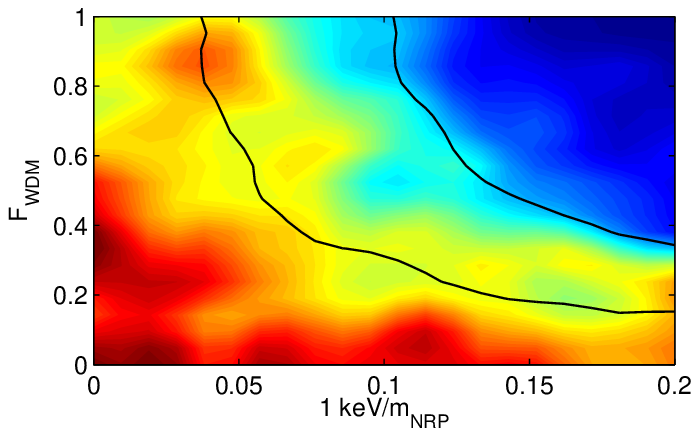}
  \caption{{\it (Left)}~Preferred region for the parameters ($\Fwdm,
    \langle v\rangle_\wdm)$ of the $\Lambda$CWDM model, using SDSS
    \lya and WMAP5 data sets. The black lines represent the $68\%$CL
    and $95\%$CL contours of the two-dimensional probability
    marginalized over other parameters (as defined in the Bayesian
    approach), while the colors show the likelihood averaged over
    other parameters.  {\it (Right)}~Same plot with the horizontal
    axis translated in terms of $(1 \mathrm{keV} / m_\dw) = \langle
    v\rangle_\wdm / (157~\mathrm{m/s})$, see equation
    (\ref{eq:vnrp}).}
  \label{fig:mwdm_fwdm}
\end{figure}

Having performed extensive tests for the case of pure $\Lambda$WDM
models, we applied the same type of analysis to the full $\Lambda$CWDM
model.  We ran {\sc CosmoMC} varying all parameters, including $0 \leq
\langle v \rangle_\wdm \leq 0.03$~km/s (corresponding to $m_\dw \ge
5\kev$) and $0\le \Fwdm\le 1$.  The resulting two-dimensional joint
probabilities are shown in Fig.~\ref{fig:cwdm} and
Fig.~\ref{fig:mwdm_fwdm}.  The data clearly prefers the pure
$\Lambda$CDM limit, which corresponds to the two axes $\Fwdm = 0$ and
$\langle v \rangle_\wdm = 0$, but significant deviations from this
model are still perfectly allowed by the data.  One can see that as
$\Fwdm$ approaches one, the $95\%$CL limit on $(\Fwdm, (1 \,
\mathrm{keV} / m_\dw))$ tends towards $(1 \mathrm{keV} / m_\dw)=0.1$,
i.e. $m_\dw=10\kev$. This bound is weaker than the $12.1\kev$
  limit obtained in the pure $\Lambda$WDM analysis, because it is
derived from a joint two-dimensional probability.\footnote{The 95\%CL
  limit in figure \vref{fig:mwdm_fwdm} contains 95\% of the posterior
  probability in the full $(\Fwdm, (1 \, \mathrm{keV} / m_\dw))$
  plane. It is clearly different to look for the restriction of this
  limit to $\Fwdm=1$, and to compute the range containing 95\% of the
  posterior probability along the $\Fwdm=1$ axis.}  There is a number
of models with $\Fwdm < 1$ and masses of the order of $5\kev$ to
$10\kev$ which lie in the region preferred by the data.  In
particular, for \Fwdm lower than 35\%, the data is compatible with the
smallest mass included in our prior range ($m_\dw=5\kev$). We did not
include smaller masses in the analysis for the reasons explained at
the end of section \ref{sec:depend-wdm-results}, but it is interesting
to see that for $m_\dw \sim (5-7)\kev$ the bound on $\Fwdm$ reaches a
flat asymptote; this suggests that for such a small WDM fraction, the
amplitude of the suppression in the small-scale flux power spectrum is
so small that the \lya data is compatible with any arbitrary value of
the free-streaming scale. In this case, considerably smaller masses
are likely to be allowed, and we would in fact need to add data on
larger scale (like galaxy redshift surveys) in order to better
constrain the free-streaming scale and the mass. This would lead to
mass limits corresponding to hot dark matter rather than warm dark
matter particles.

\subsubsection{Comparison with frequentist approach}
\label{sec:cwdm-freq}

\begin{table}[t]
  \centering
  \begin{tabular}{|l|c|c|c|}
%    & \multicolumn{5}{c|}{\fwdm} \\
    \hline
    \backslashbox{\Fwdm}{$m_\dw$} & CDM & 10\kev & 5\kev \\
     \hline 1.0 (WDM)  & 1404.0 & 1405.8 & 1415.8 \\
     \hline 0.9 & 1404.0 & 1406.4 & 1413.2 \\
     \hline 0.5 & 1404.0 & 1404.1 & 1407.4 \\
     \hline 0.2 & 1404.0 & 1404.1 & 1404.1 \\
     \hline 0.0 (CDM) & 1404.0 & 1404.0 & 1404.0 \\
    \hline
  \end{tabular}
  \caption{Values of $-\log(\CL_\text{best})$ (which play the role of $\chi^2/2$) for runs with
    fixed $m_\dw$ and \Fwdm. All runs have the same number of
    parameters, equal to that of CDM.}
  \label{tab:chi2}
\end{table}

\begin{figure}
  \centering
  \includegraphics[width=\textwidth]{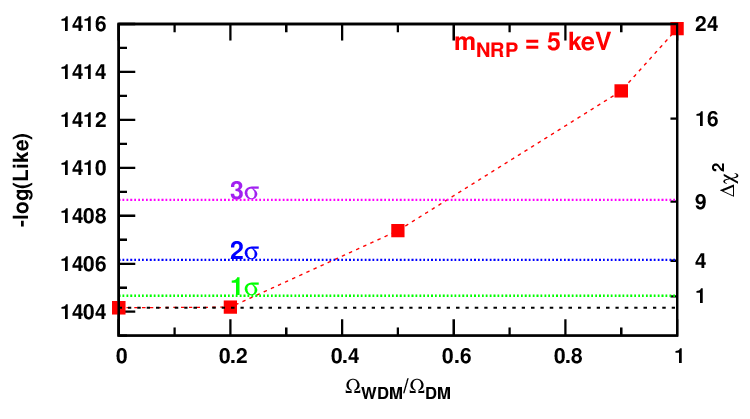}
  \caption{Dependence of $-\log(\CL_\text{best})$ on \Fwdm,
    assuming that the mass is fixed to $m_\dw=5\kev$.  Horizontal
    dashed lines show the levels $\Delta \chi^2 = 1, 4, 9$,
    corresponding to the $1,2,3\sigma$ confidence limits.}
  \label{fig:cwdm-chi2}
\end{figure}

Similarly to section~\ref{sec:runs-fixed-param} we have performed a number of
\textsc{CosmoMC} runs on a grid of points with $m_\dw=5,10 \, \kev$ and
$\Fwdm=0.2,0.5,0.9,1.0$, varying the same parameters as in the pure
$\Lambda$CDM case. For each of these runs, we report in Table~\ref{tab:chi2}
the value of $-\log(\CL_\text{best})$ in our chains, which plays the role of
$\chi^2/2$. The dependence of $\chi^2$ on \Fwdm for $m_\dw=5\kev$ is presented
in Fig.~\ref{fig:cwdm-chi2}. We see that if the mass is assumed to take this
particular value, the $2\sigma$ confidence limit ($\Delta \chi^2 =4$) becomes
$\Fwdm \simeq 40\%$, while the $2\sigma$ limit ($\Delta \chi^2 =9$) allows for
about $60\%$ of WDM. We can try to compare this case with the results of our
Bayesian analysis, for which the joint 95\%~CL limit on $(\Fwdm, (1 \,
\mathrm{keV} / m_\dw))$ reaches $\Fwdm=0.35$ in the limit $m_\dw=5$~keV. In
the frequentist approach, the joint 95\%~CL on a pair of parameters is
approximated by $\Delta \chi^2 =6.2$.  This corresponds to an upper bound
$\Fwdm \simeq 0.50$, less restrictive than the comparable Bayesian limit. For
$m_\dw = 10\kev$, all values of the WDM fraction (up to pure $\Lambda$WDM) are
allowed, in agreement with the results of section~\ref{sec:runs-fixed-param}.

\section{Systematic uncertainties}
\label{sec:syst-uncert}

The bounds derived previously assume that all systematic uncertainties
are properly taken into account in the WMAP5, VHS and SDSS \lya data
likelihoods implemented in {\sc CosmoMC}. Systematic errors are of
course hard to estimate. Here, we want to summarize the assumptions
concerning systematic errors on which our results are based, and to
describe a number of related tests.

The systematic uncertainties entering into our final mass bounds can be
roughly subdivided into:

\begin{asparaenum}[\bf 1.]

\item \emph{Observational} uncertainties related to the way the \lya forest
  data are processed to extract the flux power spectrum.  Among these
  uncertainties the most important are: $i)$ mean flux level of the data sets,
  which could be different from data set to data set and is usually kept as a
  free parameter to marginalize over in a reasonable range of observed values
  (e.g. \cite{Seljak:03}); $ii)$ the continuum of the distant QSOs, which is
  either subtracted from the data (e.g. \cite{McDonald:05}) or implicitly
  removed by considering only a smaller portion of the spectra where the
  continuum is not contributing to the flux power (e.g. \cite{Kim:2003qt});
  $iii)$ the metal absorption systems in the \lya forest, whose contribution
  to the flux power could be estimated from Voigt profile fitting statistics
  for high-resolution spectra \cite{Kim:2003qt} or modelled using the
  framework proposed by \cite{McDonald:05} for low-resolution spectra; $iv)$
  the contribution to the flux power due to the damping wings of strong damped
  \lya systems (DLAs) which again can be either removed (VHS) or modelled with
  the help of hydro simulations (SDSS); $v)$ uncertainties in the noise level
  and resolution of the instrument, which is particularly important for the
  SDSS data set.  We stress that among all these uncertainties the first two
  are the most important, while the metal line contribution and the presence
  of DLAs is modelled using parameters (constrained with priors) that are
  varied in the MCMC method. We refer to the extensive analysis made in
  Refs.~\cite{McDonald:05,McDonald:05syst} for a more quantitative discussion
  on the systematic errors and in the raw data processing technique.

\item \emph {Theoretical} uncertainties, related to \emph{numerical}
  simulations, such as the finite box size and finite numerical resolution.
  These are usually addressed by running a number of simulations that explore
  further the parameter space in resolution and box-size and checking the
  impact on the flux power. Note that these corrections are of course
  code-dependent and could be different in Eulerian and Lagrangian code. For a
  comparison between different codes we refer to \cite{Regan:06a}. It is also
  important to compare the full hydrodynamical simulations with other less
  time consuming hydro schemes such as the HPM that have been used by several
  authors. The agreement between the Eulerian code \textsc{enzo} and the
  Lagrangian code \textsc{gadget-ii} in terms of flux power is found to be
  below or around the 5\% level \cite{Regan:06a}, while the discrepancy
  between \textsc{gadget-ii} and HPM \cite{gnedinhui98} could be as high as
  20-30\% (in the low redshift bins, while at higher redshift is usually
  better and at the percent level) as found in \cite{Viel:05b}. A quantitative
  analysis of \lya flux and line properties using the code \textsc{enzo} has
  also been made in \cite{Tytler:07} with a particular focus on box-size and
  resolution problems.

\item \emph{Astrophysical} uncertainties relevant for \lya physics, e.g.
  concerning the thermal history and ionization history of the IGM. One of the
  key parameters describing the influence of the IGM on the flux power
  spectrum is $\gamma$, which appears in the empirical relation $T=T_{0}
  (1+\delta)^{\gamma-1}$ between density fluctuations and the IGM temperature.
  Recent results of~\cite{Bolton:07a} suggested an inverted
  temperature-density relation (i.e. the possibility of having $\gamma < 1$).
  In order to derive conservative bounds, it was suggested
  in~\cite{Bolton:07a} to marginalize $\gamma$ over a wide range, including
  values below $\gamma=1$.  In this case, the effect of $\gamma$ on the flux
  power spectrum is not necessarily captured by a linear approximation, and
  one should use at least a second order Taylor expansion of $P_F(k,z)$ as a
  function of $\gamma$~\cite{Bolton:07a,Viel:07}. The results of this paper
  assume a wide range and a second-order Taylor expansion for $\gamma$, as in
  these last references. This issue is particularly relevant for $\Lambda$WDM
  and $\Lambda$CWDM analyses, because the the effect of changing the
  density-temperature relation can be somewhat degenerate with that of
  introducing a cut-off in the primordial matter power spectrum. Indeed, we
  found that implementing the old method used in~\cite{Viel:05,Viel:06}
  ($\gamma >1$, first order Taylor expansion of $P_F$ in $\gamma$) affects our
  results for the mass bound $m_\dw^{95}$ by about 20\%. Since the
  density-temperature relation issue is not fully settled yet, we consider the
  uncertainty on $\gamma$ to be one important source of systematic errors in
  our analysis. Another physical ingredient which is particularly important
  and so far has been poorly modelled in hydrodynamical simulations is the
  reionization of hydrogen (which is related to the filtering scale of the
  IGM, see for example \cite{Hui:97}): estimates give a few percent impact on
  the SDSS flux power but radiative transfer effects (that are likely to
  generate temperature and ionization fluctuations at large scales) need to be
  better understood in order to quantify the effect on the flux power at
  several scales and redshifts at a level comparable to the statistical error
  of the data \cite{McDonald:04b,Croft:04}.  In principle all the physical
  processes that could impact on the flux power at large scales (roughly above
  few Mpc) need to be considered: galactic winds, star formation, cosmic rays,
  Active Galactic Nuclei (AGN) feedback, etc. (see e.g.
  \cite{Croft:2000hs,Viel:04,McDonald:05syst,Jubelgas:08}). However, we must
  stress that: $i)$ accurate investigations made with hydrodynamical
  simulations show that the impact of these effects at large scales is either
  minimal or degenerate with other parameters of the thermal history that are
  marginalized over; $ii)$ the effect of a warm dark matter component on the
  flux power spectrum does not seem to be very degenerate with other
  physical/cosmological parameters (at least in the $\Lambda$WDM case), see
  e.g. \cite{Seljak:06}.

\item Uncertainties related to the parameter extraction method.  As discussed
  already in Sections~\ref{sec:results-vhs} and~\ref{sec:depend-wdm-results},
  the ambiguity of defining a stopping criteria for the MCMCs (corresponding
  to a given degree of convergence of the chains) can lead to significant
  errors in the bounds. This error can always be reduced by increasing the
  length of the chains, but for models like $\Lambda$CWDM for which the
  execution time of {\sc camb} is larger than usual (due to the necessity of
  including several discrete momenta in the evolution equations of the warm
  component), reaching a very high degree of convergence quickly becomes
  prohibitive in terms of CPU time. The error on the bounds can be estimated
  by {\sc CosmoMC} itself, as already mention in the result sections. It
  should always be included when quoting results from {\sc CosmoMC}. For
  instance, it lead us to write our $\Lambda$WDM mass bound for the SDSS case
  as $m_{\dw}^{95}$=12.1~keV instead of 13.5~keV.

\item Finally, there are also uncertainties related to the CMB
  data. For instance, it is well-known that a different treatment of
  systematic errors in WMAP5 (beam shape, small-$l$ likelihood, etc.)
  has led to a shift of the allowed range of some parameters with
  respect to WMAP3 (e.g. $\sigma_8$). However, the results of
  Fig.~\ref{fig:cwdm} show that the parameters describing the
  warm sector are not degenerate with other cosmological parameters,
  and should therefore be rather insensitive to changes in the CMB data.
\end{asparaenum}

In summary, our estimates are based on the current state-of-the art concerning
the modelling of systematic errors involved in this field. We checked many of
these assumptions with designated test runs and found that the bounds remain
intact within $\approx 30\%$ uncertainty.  Of course, the number of
assumptions is such that derived bounds should be considered with care.  As an
illustration of this, the mass bound obtained by Seljak et
al.~\cite{Seljak:06} and Viel et al.~\cite{Viel:06} for the same model (pure
$\Lambda$WDM) and data set (SDSS \lya plus CMB data) differ by 30\% due to
variations in the treatment of systematics, data interpretation, numerical
issues, etc.

\section{Conclusion}
\label{sec:conclusion}

\begin{figure}[t]
  \centering \includegraphics[width=.8\linewidth]{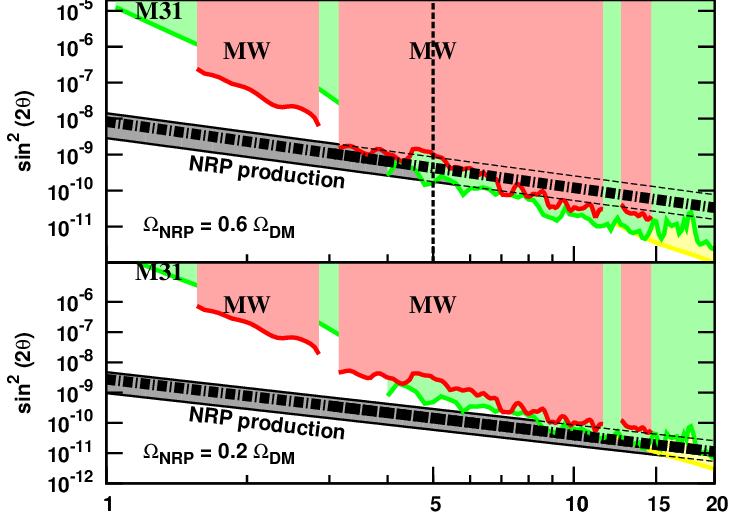}
  \caption{Comparison of X-ray bounds with the results of our \lya analysis of
    $\Lambda$CWDM, for the case in which NRP sterile neutrinos contribute to
    60\% (upper panel) or 20\% (lower panel) of the dark matter density.  The
    wide gray band (labeled ``NRP production'') shows the results of the
    computation of the NRP sterile neutrino abundance, taking into account
    uncertainties from QCD (see~\cite{Asaka:06b,Asaka:06c} for details). Â The
    X-ray bounds are from~\cite{Boyarsky:06d} (label ``MW''), from
    \cite{Boyarsky:07a} (label ``M31'') and from~\cite{Boyarsky:06c} (yellow
    region in the lower right corner). To account for possible uncertainties
    in the determination of the DM amount in various object, the X-ray bounds
    have been rescaled by a factor of 2 (c.f.~\cite{Boyarsky:07a}). The
    vertical dashed line marks the $5\kev$ lower bound on the mass found in
    the present analysis, assuming 60\% of WDM. For 20\% of DM, we found that
    the mass remains unconstrained by \lya data.}
  \label{fig:lya-xray}
\end{figure}

In this work we have performed a thorough analysis of \lya constraints
on warm and cold plus warm DM models using WMAP5 data, combined with
VHS or SDSS \lya data.  As expected from the quality of the data, the
results based on SDSS \lya data are much more constraining, and below
we only summarize this case.

Our results apply to any model in which the phase space distribution
of the warm component can be approximated by a Fermi-Dirac or rescaled
Fermi-Dirac distribution. The latter case (assuming a temperature
$T=T_\nu$) is a first-order approximation for the case of
non-resonantly produced sterile neutrinos. In most of the paper, we
expressed our mass bounds in terms of the mass $m_\dw$ defined in this
case.

Within a conservative frequentist approach, for pure $\Lambda$WDM
models, masses $m_{\dw}$ below $8\kev$ are excluded at 99.7\%\ CL
($m_\dw \ge 9.5\kev$ at 95\%\ CL).  In the case of CWDM models, our
analysis shows that for $m_\dw = 5\kev$ (the smallest WDM mass probed
in this investigation) as much as $60\%$ of WDM is allowed at
99.7\%\ CL ($40\%$ at 95\%\ CL). We believe that these results are
robust within the current state-of-the-art. Of course, as the \lya
method is still being developed, these results could be affected by
still unknown systematic uncertainties (see discussion in the
Section~\ref{sec:syst-uncert}).

From the same data we have also obtained Bayesian credible intervals on the
same parameters. The bound on $m_\dw$ (at 95\% CL), defined as in previous
works~\cite{Viel:05,Viel:06,Seljak:06}), are $m_\dw\ge 12.1\kev$ for pure
$\Lambda$WDM; for $\Lambda$CWDM, the joint probability of $(\Fwdm, 1/m_\dw)$
is displayed in figure~\ref{fig:mwdm_fwdm}: this shows that for the smallest
probed mass $5\kev$, the data is compatible with $\Fwdm \le 35\%$.

Our results can be easily translated for the case thermal relics, the
corresponding limits are the following. In the frequentist framework,
for pure $\Lambda$WDM, 
masses $m_\textsc{tr}$ below $1.5\kev$ are excluded at 99.7\%\ CL
($m_\textsc{tr} \ge 1.7\kev$ at 95\%\ CL).  In the case of CWDM models, our
analysis shows that for $m_\textsc{tr} = 1.1\kev$ (the smallest WDM mass probed
in this investigation) as much as $60\%$ of WDM is allowed at
99.7\%\ CL ($40\%$ at 95\%\ CL). In the Bayesian approach, 
for pure $\Lambda$WDM, the 95\% CL bound defined in the same way as before
reads $m_\textsc{tr} \ge 2.1 \kev$. For the $\Lambda$CWDM, the joint constraints on the mass and on the warm fraction can be easily obtained from the left
plot in figure~\ref{fig:mwdm_fwdm}, translated in terms of $m_\textsc{tr}$ using equation~(\ref{eq:vtr}).

Let us now discuss the implications of our results for models in which
the dark matter (or part of it) consists of non-resonantly produced sterile
neutrinos.  

The lower mass bound for the pure $\Lambda$WDM case should be compared
with X-ray
bounds~\cite{Boyarsky:05,Boyarsky:06b,Boyarsky:06c,Riemer:06,Watson:06,Boyarsky:06e,%
  Abazajian:06b,Boyarsky:06d,Boyarsky:06f,Boyarsky:07a}, that restrict
the mass of NRP sterile neutrinos from above. Confronting the upper
bound $m_\dw \le 4\kev$ (see e.g. Ref.~\cite{Boyarsky:07a}) with the
bound $m_\dw\ge 8\kev$ obtained in this work (both at 99.7\% CL)
excludes the NRP scenario.  Again, this conclusion should be taken
with care in view of the above discussion of systematic errors.

Next, we analyze the case of mixture of NRP sterile neutrino and some
other, cold DM particle (similar to~\cite{Palazzo:07}). To this end,
one should rescale the X-ray constraints
of~\cite{Boyarsky:05,Boyarsky:06b,Boyarsky:06c,Riemer:06,Watson:06,Boyarsky:06e,%
  Abazajian:06b,Boyarsky:06d,Boyarsky:06f,Boyarsky:07a} and the
results of the NRP DM abundance computation~\cite{Asaka:06b,Asaka:06c}
by a factor $\Fwdm<1$. We can compare the upper bounds rescaled in
this way with the results of our CWDM runs (see
Fig.~\ref{fig:lya-xray}). We see that for $\Fwdm \ge 0.6$, SDSS \lya
bounds are in conflict with X-ray constraints at $99.7\%$~CL. For an
NRP fraction $\Fwdm = 0.4 - 0.6$, an allowed window for the mass
appears around $m_\dw \approx 5\kev$ (c.f. Fig.~\ref{fig:lya-xray},
upper panel). For smaller \Fwdm, the upper mass bound from X-rays
quickly increases (as the lower panel of Fig.~\ref{fig:lya-xray}
demonstrates). At the same time, given that 1 and 2$\sigma$ \lya
contours for $\Fwdm$ become horizontal as $m_\dw$ approaches $5\kev$
(c.f.  Fig.~\ref{fig:mwdm_fwdm}), we expect that masses smaller than
5~keV with $\Fwdm \lesssim 0.4$ should be allowed (although a precise
analysis should properly take into account thermal velocities, c.f.
Sec.~\ref{sec:veloc-init-cond}). Thus, we conclude that for $\Fwdm <
0.6$ a window of allowed NRP masses opens up, and for $\Fwdm \sim 0.2$
we see that all masses (above the phase-space density
bound~\cite{Boyarsky:08a}) are allowed up to $\sim 15\kev$.

Another DM model which has similarities with the $\Lambda$CWDM scenario
discussed here is that of resonantly produced sterile
neutrinos~\cite{Shi:98,Shaposhnikov:08a,Laine:08a}.  In this case, the DM
velocity dispersion has a colder (resonant) and a warmer (NRP produced)
component. In this case, a specific analysis is needed in order to apply the
results of this work. We will present these results
elsewhere~\cite{Boyarsky:08d}.

Further improvements in the results presented here would require
performing simulations for a larger grid of points $(m_\dw,\Fwdm)$,
with a better control of systematics as well as better understanding
of the physics of the IGM. Ideally, one would like to fit all the IGM
statistics at once (flux power spectrum, flux probability distribution
function, flux bispectrum and all the line-based statistics) in the
spirit of \cite{Tytler:04,Bolton:07a} to find a consistent IGM model
from low to high redshifts.

High redshift QSOs are important for WDM (and any matter power
spectrum related) investigations, since at higher redshifts the
structures are more linear and should more closely resemble the
underlying matter power spectrum. However, at $z>5$ the IGM becomes
more neutral and the mean flux level lowers. Thus, interpreting the
almost saturated spectra would require hydrodynamical simulations at
much higher resolution in order to account for very small structures that
contribute to absorption. Furthermore, the number density of QSOs
drops significantly at high redshifts and thereby the progress will be
somewhat limited also in a statistical sense.
 
In principle, QSO spectra (especially those at high resolution and with high
signal-to-noise) carry information down to scales comparable to the resolution
element (for VLT, Very Large Telescope, UVES-spectra $\sim 0.02$ \AA\ which
could be roughly translated into 200 $h^{-1}$ comoving kpc at $z=3$), thereby
starting to probe the sub-Mpc scales. Although these numbers are very
promising, it is important to stress that at these scales the flux power
spectrum becomes very steep, so that a small change in the astrophysical and
cosmological parameters could have a large impact on it. To interpret it
with hydrodynamical simulations could become challenging both for resolution
and physical issues (all the physical ingredients that could have an influence
on galactic scales need to be modelled and their effect re-computed). It is
probably more promising to tackle some physical effects such as the IGM
thermal evolution with independent probes (like line statistics) and use them
as strong priors in the interpretation of the flux power at small scales, in
order to restrict the parameter space.  A large number of intermediate
resolution QSO spectra, between that of SDSS and VLT spectra, will be also
very important in order to cross-check the effects of systematics on the data
and improve the limits obtained in this work. This effort could be possible
quite soon with the X-shooter spectrograph \cite{Kaper:08}.

\section*{Acknowledgments}
We would like to thank M.~Laine, A.~Macci\'o, M.~Shaposhnikov for help at
various stages of this project.  We are grateful to V.~Savchenko for his
invaluable help with performing \textsc{CosmoMC} runs and adapting
\textsc{CosmoMC} to use GRID technologies.  JL acknowledges support from the
EU 6th Framework Marie Curie Research and Training network ``UniverseNet''
(MRTN-CT-2006-035863). MV thanks ASI/AAE theory grant and INFN-PD51 for
financial support. OR acknowledges support from the Swiss National Science
Foundation.
Hydrodynamical simulations were performed on the COSMOS supercomputer at DAMTP
and at High Performance Computer Cluster Darwin (HPCF) in Cambridge (UK).  The
Darwin Supercomputer of the University of Cambridge High Performance Computing
Service (http://www.hpc.cam.ac.uk) is provided by Dell Inc. using Strategic
Research Infrastructure Funding from the Higher Education Funding Council for
England.  COSMOS is a UK-CCC facility which is supported by HEFCE, PPARC and
Silicon Graphics/Cray Research.
Most MCMC simulations were performed on the computer clusters of the Ukrainian
Academic GRID, and in particular on the cluster in the Bogolybov Institute for
Theoretical Physics (Kiev) and in the Institute for Scintillation Materials
(ISMA), Kharkov. We are thankful to E.~Martynov and S.~Svistunov for providing
for us the possibility to use these facilities.
The remaining part of MCMC runs was performed on the
MUST cluster at LAPP (IN2P3/CNRS and Universit\'e de Savoie).

\appendix

\section*{Appendices}

\section{Confidence vs. credible interval}
\label{sec:confidence-credible}

The main goal of this project is to obtain reliable bounds on the
parameters of some particle physics DM models, based on the
cosmological observations of the \lya forest. This requires a careful
examination of the notion of \emph{confidence limits} in the context of
particle physics and cosmology. An extended discussion of this subject
is beyond the scope of this paper, see e.g.~\cite{MacKayBook},
or~\cite{Feldman:97}. For a short introduction see~\cite{Karlen:02}.

In particle physics one can usually repeat an experiment to determine
the value of a parameter $p$.  The measured data is a fluctuating
random variable. Then, for example, the $95\%$ \emph{confidence
  interval} on a parameter $p$ is defined as the answer to the
following question: ``What is the interval within which the parameter
$p$ will lie in $95\%$ of all measurements?''  This approach (often
called the ``frequentist'' one) is well suitable for ruling out
models.

In cosmology one needs to address a different problem, because we can
only make measurements in our observable Universe, i.e. in a single
realization of an underlying statistical model.  In the Bayesian
approach one defines, first, the likelihood $\CL(D , \{p_i\})$ of a
data set $D$ given a model with parameters $\{p_i\}$, and then the
Probability Density Function (PDF) $\Pi(\{p_i\})$ of the parameter set
$\{p_i\}$, which is equal to the likelihood $\CL(D^{obs} , \{p_i\})$
(seen as a function of the parameters, with the data fixed to the
observed ones) times the prior distribution of each
parameter~\cite{Kosowsky:2002zt,Dunkley:2004sv,Trotta:08a}.  Thus,
ultimately, the parameters (not the data!) are treated as a random
variables. For each parameter $p_i$, one can construct a probability
${\cal P}_B(p_i)$ by marginalizing (i.e., integrating) $\Pi(\{p_i\})$
over other parameters $p_{j \neq i}$, and defines the \emph{credible
  interval} as the interval of $p_i$ containing $95\%$ of the
probability ${\cal P}_B(p_i)$.

However, it is possible to estimate the frequentist \emph{confidence
  limits}, starting from the same likelihood function. 
For each parameter $p_i$, one can maximize
the likelihood over the other parameters, which gives 
an (unnormalised) probability distribution ${\cal P}_F (p_i) =
\max_{p_{j \neq i}} \CL(D^{\rm obs}, \{p_i\})$. The 95\% confidence
limits on $p_i$ are defined in such way that inside the corresponding
interval, ${\cal P}_F(p_i)$ is greater than some number
such that 95\% of possible data sets lead to a likelihood greater
than this number.

In general, the two intervals (frequentist and Bayesian) do not coincide.
However, in cosmology, most authors compute Bayesian intervals and call them
confidence limit. This is the case of all recent bounds on the WDM
mass~\cite{Viel:05,Viel:06,Seljak:06,Viel:07}.  However, in this paper, we
want to carry the Bayesian and frequentist analyses in parallel.

The Monte-Carlo Markov Chain (MCMC) method with Metropolis-Hastings
algorithm~\cite{Neal93,Lewis:02,Kosowsky:2002zt,Trotta:08a},
implemented in the code {\sc CosmoMC}~\cite{Lewis:02}, became a
standard method for exploring the Bayesian PDF $\Pi(\{p_i\})$, which
is equal to the likelihood $\CL(D^{obs} , \{p_i\})$ (modulo a
normalization factor) in case of flat priors.
The MCMC method is very effective in scanning $\Pi$ (or
$\CL$) in the region where this function is the largest,
with the crucial property that once the chains have converged, the density of
points in parameter space is proportional to $\Pi(\{p_i\})$.  This
makes it an ideal tool for a Bayesian statistical analysis (see
e.g.~\cite{Trotta:08a} and references therein), since the credible
interval for a given parameter marginalized over other
parameters is simply obtained by counting the number of chain points in each
parameter bin.

In order to obtain frequentist confidence limits on a parameter $p_i$, we
should take into account the exact form of the likelihood as a function of the
data (in cosmology, the likelihood is usually non-Gaussian, in particular for
small-scale CMB data).  Instead, we choose to use the same approximation as in
many frequentist analyses: we define a $\chi^2$ function as
$\chi^2(p_i)=\min_{p_{j \neq i}} [-2 \ln \CL]$, search for its minimum
$\chi^2_{\min}=\min_{p_i} [\chi^2(p_i)]$, and finally define the
$1\sigma,2\sigma, 3\sigma$ confidence intervals as the region in which $\Delta
\chi^2(p_i) \equiv \chi^2(p_i) - \chi_{\min}^2$ is smaller than $1,4$ and $9$.
In order to minimize $[-2 \ln \CL]$ over $p_{j \neq i}$, we still use {\sc
  CosmoMC} with $p_i$ kept fixed to different values.  It is well--known that
the {\sc CosmoMC} algorithm is not optimal for minimization (with respect,
e.g., to a simulated annealing method), but in this work it turned out to be
efficient enough.

%\bibliographystyle{JHEP-2}%
%\bibliography{preamble,astro,dima,cosmomc,jl}
\let\jnlstyle=\rm\def\jref#1{{\jnlstyle#1}}\def\aj{\jref{AJ}}
  \def\araa{\jref{ARA\&A}} \def\apj{\jref{ApJ}\ } \def\apjl{\jref{ApJ}}
  \def\apjs{\jref{ApJS}} \def\ao{\jref{Appl.~Opt.}} \def\apss{\jref{Ap\&SS}}
  \def\aap{\jref{A\&A}} \def\aapr{\jref{A\&A~Rev.}} \def\aaps{\jref{A\&AS}}
  \def\azh{\jref{AZh}} \def\baas{\jref{BAAS}} \def\jrasc{\jref{JRASC}}
  \def\memras{\jref{MmRAS}} \def\mnras{\jref{MNRAS}}
  \def\pra{\jref{Phys.~Rev.~A}} \def\prb{\jref{Phys.~Rev.~B}}
  \def\prc{\jref{Phys.~Rev.~C}} \def\prd{\jref{Phys.~Rev.~D}}
  \def\pre{\jref{Phys.~Rev.~E}} \def\prl{\jref{Phys.~Rev.~Lett.}}
  \def\pasp{\jref{PASP}} \def\pasj{\jref{PASJ}} \def\qjras{\jref{QJRAS}}
  \def\skytel{\jref{S\&T}} \def\solphys{\jref{Sol.~Phys.}}
  \def\sovast{\jref{Soviet~Ast.}} \def\ssr{\jref{Space~Sci.~Rev.}}
  \def\zap{\jref{ZAp}} \def\nat{\jref{Nature}} \def\iaucirc{\jref{IAU~Circ.}}
  \def\aplett{\jref{Astrophys.~Lett.}}
  \def\apspr{\jref{Astrophys.~Space~Phys.~Res.}}
  \def\bain{\jref{Bull.~Astron.~Inst.~Netherlands}}
  \def\fcp{\jref{Fund.~Cosmic~Phys.}} \def\gca{\jref{Geochim.~Cosmochim.~Acta}}
  \def\grl{\jref{Geophys.~Res.~Lett.}} \def\jcp{\jref{J.~Chem.~Phys.}}
  \def\jgr{\jref{J.~Geophys.~Res.}}
  \def\jqsrt{\jref{J.~Quant.~Spec.~Radiat.~Transf.}}
  \def\memsai{\jref{Mem.~Soc.~Astron.~Italiana}}
  \def\nphysa{\jref{Nucl.~Phys.~A}} \def\physrep{\jref{Phys.~Rep.}}
  \def\physscr{\jref{Phys.~Scr}} \def\planss{\jref{Planet.~Space~Sci.}}
  \def\procspie{\jref{Proc.~SPIE}} \let\astap=\aap \let\apjlett=\apjl
  \let\apjsupp=\apjs \let\applopt=\ao
\providecommand{\href}[2]{#2}\begingroup\raggedright\endgroup

\end{document}